\documentclass[9pt,twocolumn,twoside]{pnas-new}

\templatetype{pnasresearcharticle} 

\makeatletter
\@ifundefined{textcolor}{}
{%
 \definecolor{BLACK}{gray}{0}
 \definecolor{WHITE}{gray}{1}
 \definecolor{RED}{rgb}{1,0,0}
 \definecolor{GREEN}{rgb}{0,1,0}
 \definecolor{BLUE}{rgb}{0,0,1}
 \definecolor{CYAN}{cmyk}{1,0,0,0}
 \definecolor{MAGENTA}{cmyk}{0,1,0,0}
 \definecolor{YELLOW}{cmyk}{0,0,1,0}
}

 \@ifundefined{textcolor}{}{%
  \definecolor{BLACK}{gray}{0}
  \definecolor{WHITE}{gray}{1}
  \definecolor{RED}{rgb}{1,0,0}
  \definecolor{GREEN}{rgb}{0,1,0}
  \definecolor{BLUE}{rgb}{0,0,1}
  \definecolor{CYAN}{cmyk}{1,0,0,0}
  \definecolor{MAGENTA}{cmyk}{0,1,0,0}
  \definecolor{YELLOW}{cmyk}{0,0,1,0}
 }

 \@ifundefined{textcolor}{}{
  \definecolor{BLACK}{gray}{0}
  \definecolor{WHITE}{gray}{1}
  \definecolor{RED}{rgb}{1,0,0}
  \definecolor{GREEN}{rgb}{0,1,0}
  \definecolor{BLUE}{rgb}{0,0,1}
  \definecolor{CYAN}{cmyk}{1,0,0,0}
  \definecolor{MAGENTA}{cmyk}{0,1,0,0}
  \definecolor{YELLOW}{cmyk}{0,0,1,0}
 }
 \@ifundefined{definecolor}{
 }{}
 \@ifundefined{definecolor}{color}{}

\newcommand{\be}{\begin{equation}}
\newcommand{\ee}{\end{equation}}
\newcommand{\bea}{\begin{eqnarray}}
\newcommand{\eea}{\end{eqnarray}}
\newcommand{\bse}{\begin{subequations}}
\newcommand{\ese}{\end{subequations}}

\definecolor{d_red}{cmyk}{0.00, 0.81, 1.00, 0.27}
\definecolor{d_orange}{cmyk}{0.00, 0.33, 1.00, 0.00}
\definecolor{d_blue}{cmyk}{0.78, 0.47, 0.00, 0.20}
\definecolor{d_lgreen}{cmyk}{0.07, 0.00, 0.79, 0.29}
\definecolor{d_green}{cmyk}{0.66, 0.00, 0.71, 0.56}
\definecolor{d_blue}{cmyk}{0.78, 0.47, 0.00, 0.20}
\definecolor{d_dblue}{cmyk}{0.91, 0.79, 0.00, 0.22}
\definecolor{d_pink}{cmyk}{0.0, 0.79, 0.37, 0.29}
\definecolor{d_purple}{cmyk}{0.16, 0.54, 0.00, 0.70}
\definecolor{d_paleblue}{cmyk}{0.669, 0.338, 0.00, 0.373}
\definecolor{d_dpaleblue}{cmyk}{0.441, 0.290, 0.00, 0.580}
\definecolor{d_brown}{cmyk}{0.0, 0.490, 0.930, 0.350}
\definecolor{d_turquoise}{cmyk}{0.630, 0.04, 0.0, 0.440}
\definecolor{KIT-green}{RGB}{0, 150,130}
\definecolor{KIT-blue}{RGB}{70,100,170}

\def\bmx{\begin{pmatrix}}
\def\emx{\end{pmatrix}}

\makeatother

\title{High sensitivity heat capacity measurements on {\boldmath Sr$_2$RuO$_4$} under uniaxial pressure}

\author[a,b]{You-Sheng Li}
\author[c]{Naoki Kikugawa}
\author[a]{Dmitry A.\ Sokolov }
\author[a]{Fabian Jerzembeck}
\author[d]{Alexandra S.\ Gibbs}
\author[e]{Yoshiteru Maeno }
\author[a]{Clifford  W.\ Hicks}
\author[f,g]{J\"{o}rg Schmalian}
\author[a,1]{Michael Nicklas}
\author[a,b,1]{Andrew P.\ Mackenzie}

\affil[a]{Max Planck Institute for Chemical Physics of Solids, N\"othnitzer Str.~40, 01187 Dresden, Germany}
\affil[b]{Scottish Universities Physics Alliance, School of Physics and Astronomy, University of St Andrews, St Andrews KY16 9SS, UK}
\affil[c]{National Institute for Materials Science, Tsukuba 305-0003, Japan}
\affil[d]{ISIS facility, STFC Rutherford Appleton Laboratory, Chilton, Didcot, OX11 0QX, UK}
\affil[e]{Department of Physics, Graduate School of Science, Kyoto University, Kyoto, Japan}
\affil[f]{Institut f\"{u}r Theorie der Kondensierten Materie, Karlsruher Institut f\"{u}r Technologie, 76131 Karlsruhe, Germany}
\affil[f]{Institut f\"{u}r Quantenmaterialien und -technologien, Karlsruher Institut f\"{u}r Technologie, 76131 Karlsruhe, Germany}

\leadauthor{Li}

\significancestatement{Research on the unconventional superconductivity of Sr$_2$RuO$_4$ is undergoing a renaissance since recent spin susceptibility measurements ruling out the spin triplet order parameter which had been widely favored for over two decades.  With ultrasound, Kerr rotation and muon spin relaxation data all providing evidence for a two-component order parameter, it is vital that this possibility be investigated thermodynamically by studying the dependence of the heat capacity anomaly on uniaxial pressure.  Here, the relevant experimental results are combined with theoretical analysis that shows how strongly the data constrain theories of the order parameter. In particular, we do not observe any signs of transition splitting of two order parameter components. Sr$_2$RuO$_4$ thus offers a unique test-bed for theories of unconventional superconductivity. }

\authordeclaration{The authors declare no conflict of interest.}

\authordeclaration{Data deposition: The data that underpin the ﬁndings of this study are available from the
University of St Andrews research portal: https://doi.org/10.17630/b5ff3cfd-bb2d-46c9-a240-7a7064562588.}

\correspondingauthor{\textsuperscript{1}To whom correspondence should be addressed. E-mail: Michael.Nicklas@cpfs.mpg.de or Andy.Mackenzie@cpfs.mpg.de}

\begin{abstract}
A key question regarding the unconventional superconductivity of {\boldmath Sr$_2$RuO$_4$} remains whether the order parameter is single- or two-component.  Under a hypothesis of two-component superconductivity, uniaxial pressure is expected to lift their degeneracy, resulting in a split transition.  The most direct and fundamental probe of a split transition is heat capacity.  Here, we report measurement of heat capacity of samples subject to large and highly homogeneous uniaxial pressure.  We place an upper limit on the heat-capacity signature of any second transition of a few per cent of that of the primary superconducting transition. The normalized jump in heat capacity, {\boldmath $\Delta C/C$}, grows smoothly as a function of uniaxial pressure, favouring order parameters which are allowed to maximize in the same part of the Brillouin zone as the well-studied van Hove singularity.
Thanks to the high precision of our measurements, these findings place stringent constraints on theories of the superconductivity of {\boldmath Sr$_2$RuO$_4$}.
\end{abstract}

\dates{This manuscript was compiled on \today}

\begin{document}

\verticaladjustment{-2pt}

\maketitle
\thispagestyle{firststyle}
\ifthenelse{\boolean{shortarticle}}{\ifthenelse{\boolean{singlecolumn}}{\abscontentformatted}{\abscontent}}{}

\dropcap{O}btaining a full understanding of the superconductivity of Sr$_2$RuO$_4$ is a core challenge for condensed matter physics.  Since soon after its discovery over a quarter of a century ago \cite{Maeno1994}, the superconducting order parameter of Sr$_2$RuO$_4$ has been known to be unconventional \cite{Mackenzie2003,Mackenzie1998}, and to condense from a well-understood and fairly simple quasi-two-dimensional Fermi liquid metallic state \cite{Mackenzie1996,Maeno1997,Bergemann2000,Bergemann2003}.  Given the profound advances in theoretical techniques in recent decades a full understanding of its superconductivity is an important, and attainable, challenge for the field.  The form of the wave-vector dependent susceptibility of Sr$_2$RuO$_4$ leads to the prediction of a rich superconducting phase diagram in weak-coupling calculations which aim to perform a bias-free estimate of the condensation energies of different order parameters.  A notable feature of the results is how close a number of different odd- and even-parity solutions are seen to be in energy \cite{Raghu2010,Scaffidi2014,Roising2019}.  On the one hand this emphasizes the potential of Sr$_2$RuO$_4$ as a test-bed material on which to refine the predictive capabilities of modern theories of unconventional superconductivity \cite{Mackenzie2017}.  On the other hand, realising this potential will likely first require a conclusive experimental determination of which of the many possible order parameters wins out in the real material.  This is a particularly exciting stage of the quest to complete this empirical determination, for reasons that we will now outline.

For over twenty years, the large majority of attention was paid to odd-parity order parameter candidates for Sr$_2$RuO$_4$ \cite{Maeno2012}, because of nuclear magnetic resonance (NMR) measurements of spin susceptibility in the superconducting state that seemed to be inconsistent with any even-parity order parameter \cite{Burganov2016}.  However, thanks to the discovery of a systematic error in the original NMR measurements \cite{Pustogow2019,Ishida2019}, that situation has now been more or less completely reversed.  Taking into account the most recent measurements of the magnetic field dependence of the spin susceptibility \cite{Chronister2020}, it seems clear that the order parameter must be even parity or at least dominated by an even parity component.  The spin susceptibility results would be most easily describable in terms of a single-component, likely $d$-wave, order parameter, but recent thermodynamic evidence from ultrasound experiments is most straightforwardly interpreted in terms of an order parameter with two degenerate components \cite{Benhabib2020,Ghosh2020}.  Such order parameters do not of necessity break time-reversal symmetry, but they can, if the two degenerate have the appropriate phase relationship.  In the context it is significant that long-standing muon-spin relaxation ($\mu$SR) \cite{Luke1998,Luke2000} and magneto-optic Kerr rotation measurements \cite{Xia2006} have indicated time-reversal symmetry breaking in the superconducting state.

To investigate any order parameter with two degenerate components, whether or not it breaks time-reversal symmetry, uniaxial pressure is a powerful probe because it can split the degeneracy, creating a split superconducting phase transition \cite{Sigrist1991}.  In a significant experimental advance, the muon-spin relaxation experiments have recently been extended to high uniaxial pressures \cite{Grinenko2020}.  In line with naive expectation, the temperature at which time-reversal symmetry is broken ($T_{TRSB}$) splits from the main superconducting transition ($T_c$), with $T_{TRSB}$ remaining nearly pressure-independent while $T_c$ increases under the application of the pressure.  However, there has been a long-standing question about whether the Kerr and muon signals correspond to bulk thermodynamic transitions, so it is highly desirable to compare the new muon-spin relaxation data with those from a bulk thermodynamic probe.
In this context, it is natural to look at heat capacity, because it has an intrinsic sensitivity to transitions within the superconducting state, as is well known from work on UPt$_3$ \cite{Fisher1989,Hasselbach1989}.

\section*{Experiment}
Both for the above reasons and to give the now widely applied uniaxial pressure measurements a solid thermodynamic foundation, we have developed a high-frequency ac technique for measuring heat capacity in a uniaxial pressure cell. Full technical details are given in \cite{Li2020}; here we summarise the governing relationship of the measurement:
\begin{equation}
  C_{ac}=  \frac{P}{\omega T_{ac} } F(\omega),
\label{Cac}
\end{equation}
where $T_{ac}$ is the measured amplitude of the temperature oscillation in response to an oscillatory heat input, $P$ is the power associated with that heat input, and $\omega$ is the angular frequency.  $F(\omega)$ is the frequency response curve that characterizes the thermalization of the sample, and depends on the time constants, thermal conductances and heat capacities of the system. In the very high frequency version of the technique that we employ to restrict the probed volume to the most homogeneously strained central portion of the sample, the probed volume depends on the sample’s thermal conductivity.  The quantity $P/(\omega T_{ac})$ is therefore closely related, but not identical, to the intrinsic heat capacity of the sample. This is not a major restriction because the sensitivity of the measurement to phase transitions is hardly affected, and robust analysis of important quantities such as the ratio of the heat capacity jump to the normal state heat capacity can still be carried out.  However, to emphasize that these are not fully calibrated measurements of the full sample heat capacity, we begin by showing unprocessed experimental data.

\section*{Results}

\begin{figure}[tb!]
\centering
\includegraphics[width=0.9\linewidth]{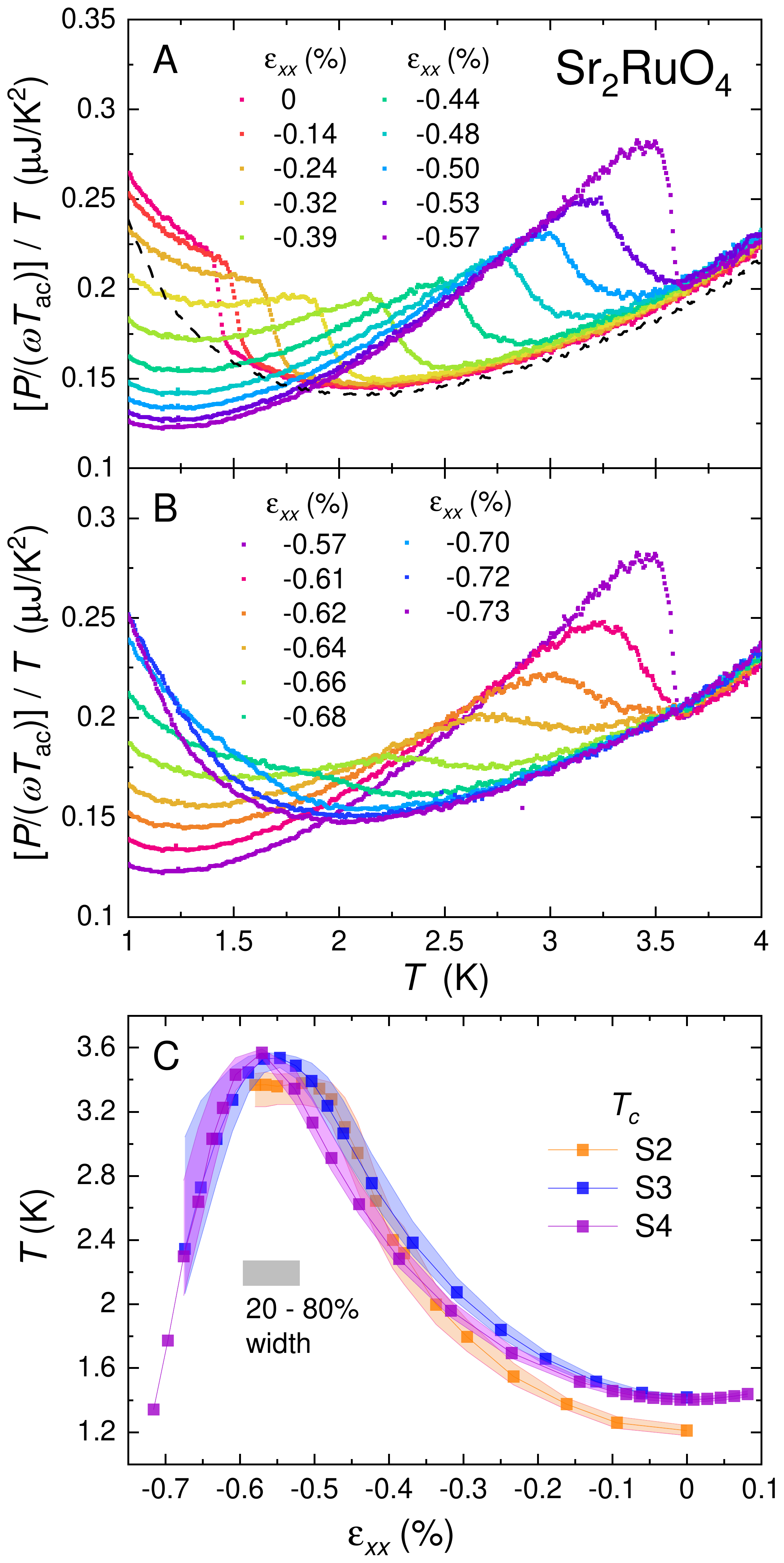}
\caption{
(\textit{A}) Experimental heat capacity data for strains between zero and $\varepsilon_{xx}= -0.57\%$ and (\textit{B}) for $|\varepsilon_{xx}|\geq 0.57\%$.  $P$ is the applied heater power and $T_{ac}$ the temperature oscillation amplitude in response; for further explanation, see the main text. The dashed line in panel (\textit{A}) is a measurement at $\mu_0H_{\parallel c} = 0.1$~T and $\varepsilon_{xx}= 0$.  (\textit{C}) The transition temperatures as a function of strain for three different samples, extracted from the leading edges of the superconducting anomalies, based on data taken at similar frequencies (3.5~kHz for sample S2 and 3.9~kHz for samples S3, S4).  The solid points are based on the midpoints ($50\%$) of the leading edge, and the shading represents the breadth of the transition from the $20\%$ to $80\%$ levels.
}
\label{HC}
\end{figure}

In Figs.~\ref{HC}\textit{A} and \ref{HC}\textit{B} we display raw data for $P/(\omega T_{ac})$ as a function of temperature for strains $\varepsilon_{xx}$ applied along the crystallographic $[100]$ directionbelow and above that at which $T_c$ of Sr$_2$RuO$_4$ is maximized.  A background signal exists, but this can straightforwardly be identified by the application of a magnetic field of 0.1~T at zero applied pressure (dashed black line in Fig.~\ref{HC}\textit{A}). Data toward lower temperatures can be found in the \textit{SI Appendix, Heat Capacity Measurements in the Temperature Range between 0.5 and 4~K}. That it is essentially independent of the strain in the sample is demonstrated by how closely it is followed at all temperatures above $T_c$, for all strains (Figs.~\ref{HC}\textit{A} and \ref{HC}\textit{B}).  The heat capacity anomaly associated with superconductivity is clearly visible in the raw data, and identifying $T_c$ from the leading edge of the anomaly allows its strain dependence to be plotted in Fig.~\ref{HC}\textit{C}.  It is seen to be in excellent agreement with previous determinations based on magnetic susceptibility \cite{Hicks2014b,Steppke2017}; the current data fully confirm that it is a bulk phenomenon.  Although we refer to \cite{Li2020} for the details of all the experimental steps necessary to obtain data of the quality shown in Fig.~\ref{HC}, we note in passing that the required specifications were demanding, and in particular that low temperature amplification was employed to reduce the voltage noise on the thermocouple reading used to determine $T_{ac}$ to less than 20~pVHz$^{-1/2}$.

\begin{figure}[tb!]
\includegraphics[width=0.9\linewidth]{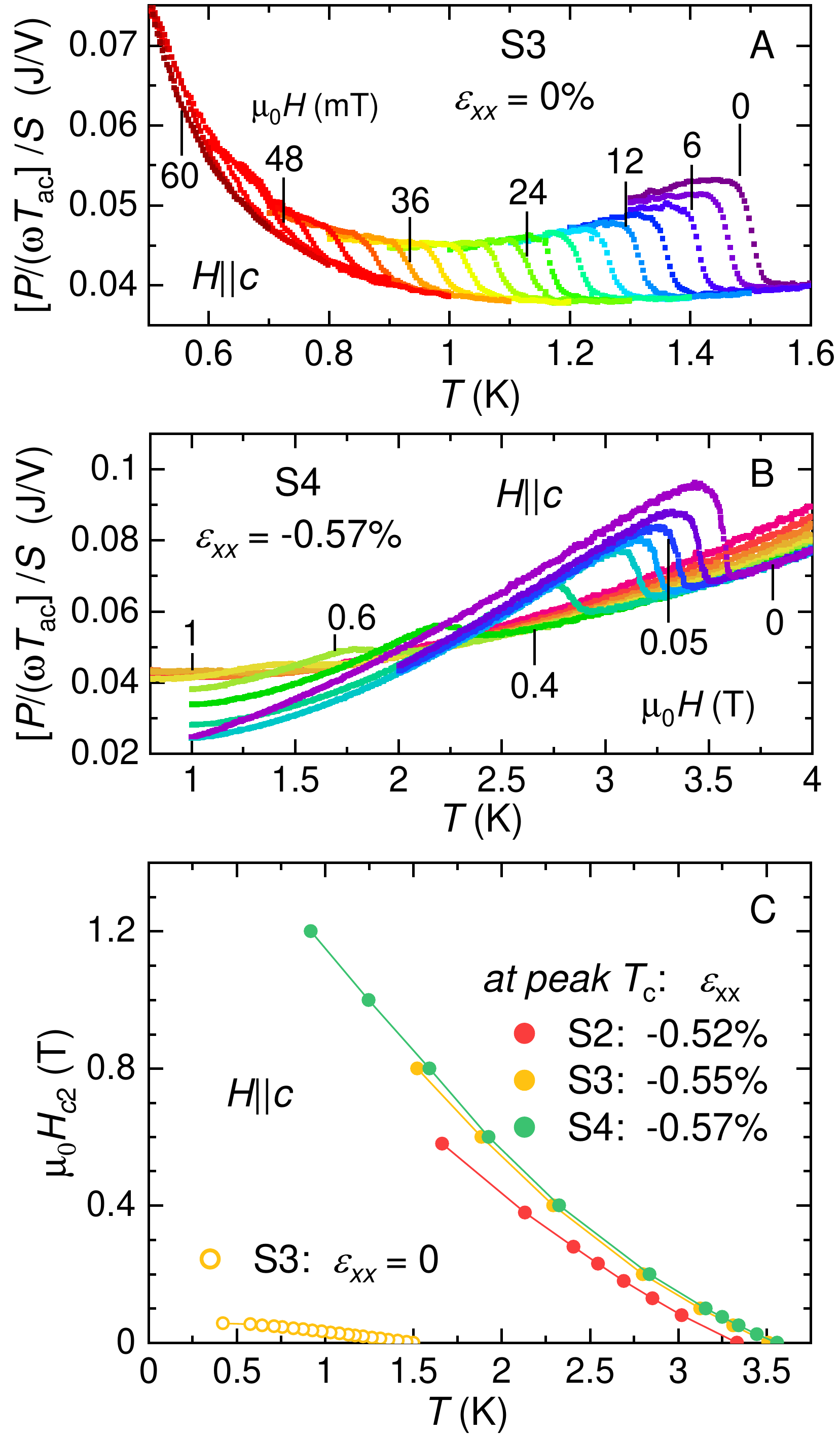}
\centering
\caption{
(\textit{A})Experimental heat capacity data for magnetic fields  $\mu_0H$ between zero and 60~mT at $\varepsilon_{xx}= 0$ and (\textit{B}) $0 \leq \mu_0H \leq 1$~T for  $-0.57\%$.  For explanation of the choice of plotted $y$-axis quantity see main text. (\textit{C}) Critical fields as a function of temperature for sample S3 under zero strain and at its peak $T_c$, and for samples S2 and S4 at their peak values of $T_c$.
}
\label{HC_field}
\end{figure}

It was also possible to study the magnetic field dependence of the heat capacity anomaly with high precision, as is demonstrated in Figs.~\ref{HC_field}\textit{A} and \textit{B} for unstrained and $T_c$-maximized samples respectively. We note that in Fig.~\ref{HC_field} $P/(\omega T_{ac})$ is normalized by the Seebeck coefficient since the thermocouple is not calibrated in field. This does not have any influence on the position of the anomaly. The anomaly remains well-resolved over wide ranges of field, enabling the critical field curves to be deduced with confidence. Again, these are in good agreement with results previously obtained from magnetic susceptibility \cite{Steppke2017}, and in particular confirm the slightly surprising observation that the shape of the critical field curves changes from the standard ‘convex’ curvature in unstrained material to a concave curvature when strained to the maximum $T_c$.  The heat capacity measurements confirm that this is a bulk effect, and the data invite further theoretical attention.

\begin{figure}[tb!]
\includegraphics[width=0.9\linewidth]{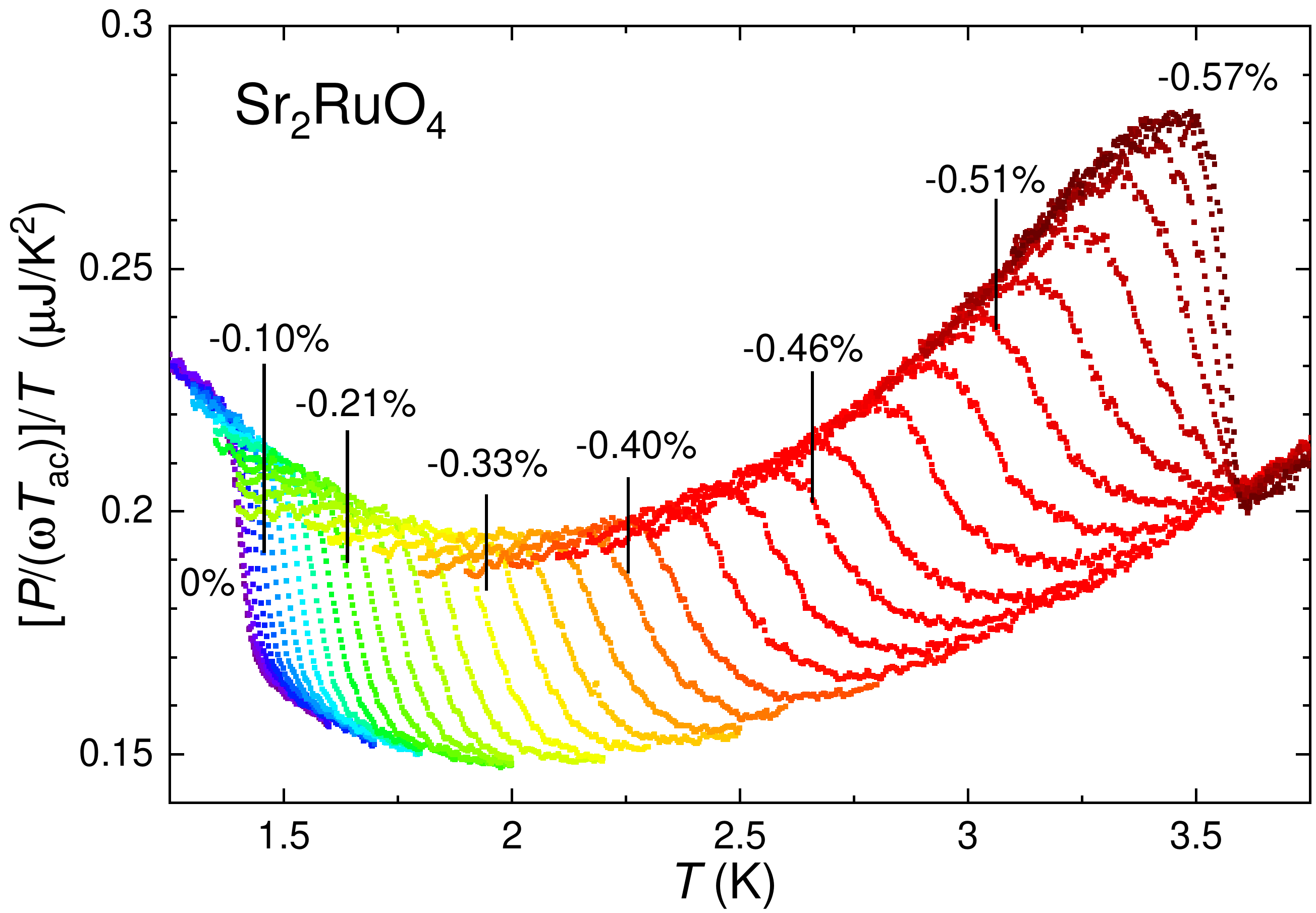}
\centering
\caption{
(Experimental heat capacity data for sample S4 for thirty-six different strains between $\varepsilon_{xx}= 0$ and $\varepsilon_{xx}= -0.57\%$.  For explanation of the choice of plotted $y$-axis quantity see main text.
}
\label{HC_smallsteps}
\end{figure}

As stated above, one of the naive expectations for two-component superconducting order parameters subject to uniaxial pressure is transition splitting.  Heat capacity is a much more direct probe of such splitting than resistivity or magnetic susceptibility, because transitions in the superconducting state are not shorted or screened by the onset of the higher temperature superconducting phase.  It is therefore particularly significant that there is no convincing qualitative evidence for such a splitting in our data, either as a function of strain or of magnetic field.  We have checked this carefully with two further measurements.  Firstly, we have carefully tracked the evolution of the leading edge of the anomaly with strain by sweeping the temperature through the transition at thirty-six distinct fixed strains, with the results displayed in Fig.~\ref{HC_smallsteps}.  There is some transition broadening at intermediate strains, which can be understood because we have approximately $10\%$ strain inhomogeneity across the probed portion of the sample, which broadens the observed transitions for the values of $\varepsilon_{xx}$ for which $dT_c/d\varepsilon_{xx}$ is large.  We show all the actual data so that the readers can judge for themselves whether there is any clear evidence for splitting beyond this broadening; in our judgement there is none.

The second check that we have performed is explicitly informed by the findings reported in \cite{Grinenko2020}, in which muon-spin relaxation data indicate that the onset of time reversal symmetry breaking at $T_{TRSB}$ occurs in the range $1.2 - 1.5$~K for all strains lower than that at which $T_c$ is maximized.  By multiple averaging of data from the temperature range of interest, we reduced the r.m.s.\ voltage noise on the thermocouple signal to 0.5~pV, giving a detection limit for a sharp secondary transition of only $0.3\%$ of the size of the primary one or, more realistically, a detection limit of $5\%$ for a secondary transition of similar width to the that of the primary one (see \textit{SI Appendix, Experimental Limits} for full details).  Even the weaker $5\%$ limit has strong implications for the underlying physics of Sr$_2$RuO$_4$, as we discuss below.

\begin{figure}[tb!]
\includegraphics[width=0.9\linewidth]{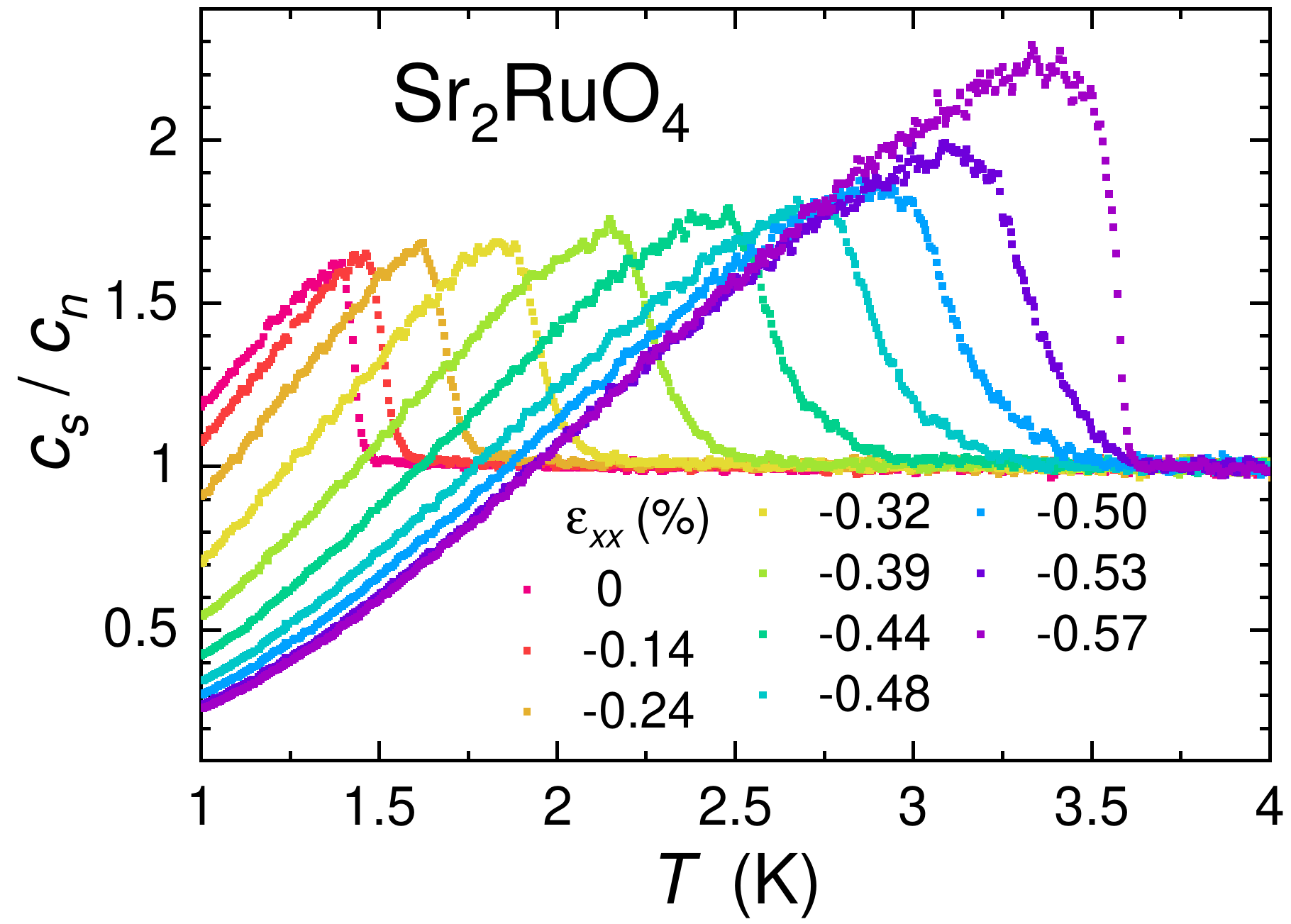}
\centering
\caption{
Superconducting state heat capacity normalized to the normal-state value, showing the evolution of the anomaly height $\Delta c_s/c_n$ with $T_c$.  Data for the relative anomaly height are accurate, while the data below $T_c$ are subject to small systematic errors described in the \textit{SI Appendix, Heat Capacity Jump} (see also main text).
}
\label{CnCs}
\end{figure}

Before moving on to that discussion, we present our strain-dependent data after one step of processing, in order to gain further physical insight.  Our measurement is not absolute, for the reasons described above, but it is possible to extract the ratio of the superconducting heat capacity to that in the normal state using a procedure (described in the \textit{SI Appendix, Heat Capacity Jump}) that gives an accurate estimate of the normalized heat-capacity jump at $T_c$ ($\Delta c_s/c_n$, where $c_s$ and $c_n$ are the heat capacities in the superconducting and normal states respectively) and is subject to error only at the level of tens of per cent for $0.7T_c\leq T < T_c$.  The result of applying this analysis to data from a representative sample of Sr$_2$RuO$_4$ is shown in Fig.~\ref{CnCs}.  Importantly, $\Delta c_s/c_n$ is seen to grow as the uniaxial pressure increases the superconducting transition temperature.  The significance of this observation is the following.  There is now good evidence from calculations \cite{Steppke2017,Hsu2016,Barber2019b} and direct spectroscopic measurement \cite{Sunko2019} that the microscopic process that maximizes $T_c$ in Sr$_2$RuO$_4$ is tuning the Fermi level through a van Hove singularity, creating a pronounced maximum in the density of states.  Within the two-dimensional Brillouin zone, the density of states is maximized near the zone boundary at the $M$ point, so for any order parameter for which the gap has a symmetry-imposed zero at this point, $\Delta c_s/c_n$ would naively be expected to fall, even as $T_c$ was maximized.  We observe the reverse effect, so our data are consistent with order parameters whose gap is allowed by symmetry to be maximal in this region of the Brillouin zone.

\section*{Discussion}

We close by discussing the relationship between our results and those obtained from other experiments, and the significance of our findings for the development of theory.  The ultrasound measurements of \cite{Benhabib2020,Ghosh2020} probe another thermodynamic quantity, so one would expect consistency between the conclusions drawn there and the current findings.  That is the case: the discontinuities observed in ultrasound velocity are consistent with transition splitting for pressure applied along the $[110]$ direction, but no discontinuity was resolved in $[100]$ modes. If the relevant physics is dominated by effects that are linear in strain couplings, the lack of splitting seen here as a function of $[100]$ pressure is not in contradiction with those experiments.  Future heat capacity work with pressure applied along $[110]$ is desirable; in \cite{Ghosh2020} the splitting $T_c - T_{TRSB}$ is estimated to grow at a rate of at least $2~{\rm K}/\%$-strain. In principle this could be resolved using our techniques.

In contrast to the ultrasound experiments, $\mu$SR does observe a split between the onset of spontaneous magnetism at $T_{TRSB}$ and that of superconductivity at $T_c$ in samples subject to $[100]$ uniaxial pressure \cite{Grinenko2020}. One possibility for the apparent discrepancy between the two experiments is the inability of ultrasound measurements performed at MHz to resolve an attenuation discontinuity against the very large background attenuation seen in the $[100]$ channel. This consideration would be particularly relevant if the thermodynamic signature associated with the muon signal were very small.  Our measurements confirm empirically that the latter is true: no heat capacity anomaly is observed at $T_{TRSB}$ within conservative detection limits of $5\%$ of the size of the anomaly at $T_c$. However, that does not remove the key challenge of reconciling such a small second anomaly with a TRSB state, for reasons we now outline.

The possible pairing states of the unstrained system
can be naturally grouped into three categories: i) $E$-pairing, which
is two-component and allows, alternatively, for time-reversal breaking $E$-pairing, such as $d_{xz}\pm id_{yz}$,
for $B_{2g}$-nematic $E$-pairing $d_{xz}\pm d_{yz}$, or for $B_{1g}$-nematic
$E$-pairing $d_{xz}$ or $d_{yz}$, ii) two accidentally degenerate
single component pairing states, such as $d\pm ig$ or $d\pm is$,
and  iii) single component pairing such as the $d_{x^{2}-y^{2}}$ state. While none of these states naturally accounts for all observations, our results offer strong evidence against some and give guidance which future experiments will allow to further discriminate between the remaining options for superconducting order.

 A key finding of the $\mu$SR experiment is that the strain dependence of $T_{TRSB}$ is much weaker than that of $T_c$.  If we accept that odd parity order parameters are ruled out by the recent spin susceptibility experiments, the only even-parity state with a symmetry-protected degeneracy at $T_c$, i.e. the only even-parity candidate for non-accidental TRS breaking at $T_c$ in unstrained Sr$_2$RuO$_4$, is $E_g$-pairing of the form $d_{xz} + id_{yz}$ \cite{Zutic2005,suh2019}.  However, to leading order in strain, in Ginzburg-Landau theory such a state obeys the relationship that when the transition is split under uniaxial pressure, the ratio of the heat capacity anomaly jumps at the two transitions is inversely proportional to the strain dependences of the transition temperatures (for the derivation see the \textit{SI Appendix, Heat Capacity Anomaly with $E_{g}$ or $E_{u}$ Pairing under Strain}).  For Sr$_2$RuO$_4$, this would imply a larger heat capacity anomaly at $T_{TRSB}$ than at $T_c$, in sharp contrast to our observations.

In Ref.\ \cite{Grinenko2020}, it was argued that
the discrepancy between heat capacity ratios and $T_c$ slopes could be explained by higher-order terms that dominate the leading order terms in the Ginzburg-Landau expansion even at relatively low strains. Here, we argue that although this is a reasonable hypothesis in the vicinity of the Lifshitz transition, where the evolution of the electronic structure is clearly nonlinear in strain, it appears less likely at low strains, and therefore that the absence of a resolvable second heat capacity anomaly makes TRSB $E$-pairing $d_{xz} \pm id_{yz}$ order unlikely. The same is true for $B_{2g}$-nematic $E$-pairing with the additional problem that it does not account for time-reversal symmetry breaking.

For even parity superconductivity to be consistent with both the heat capacity and  $\mu$SR  it would need to involve some special tuning to yield the required exceptionally small heat capacity anomaly at $T_{TRSB}$. That naturally invites examination of states in which the degeneracy of the two components at $T_c$ is accidental, and this relationship does not hold.  A recently-proposed $d + ig$ state \cite{Kivelson2020} has the attractive feature that it predicts a pattern of ultrasound anomalies that is in agreement with experiment.  However, the limit we place here on the size of the anomaly at $T_{TRSB}$ is hard to reproduce (as the authors of \cite{Kivelson2020} discuss).  The problem of predicting a heat capacity anomaly at $T_{TRSB}$ that is too large to be consistent with our experiments is even more pronounced for theories involving other accidentally degenerate combinations of even parity states such as $d+is$ \cite{Romer2019}, because unlike for the proposed $d+ig$ state, the $s$ component does not have a node in the same place as the $d$ component.  Intriguingly, a different postulate of a mixed parity order parameter with majority even and minority odd components \cite{Scaffidi2020} solves the problem of the lack of heat capacity anomaly at $T_{TRSB}$ in a rather natural way.  However, the predictions of that theory do not match the pattern of ultrasound anomalies and the theory is also strongly constrained by the newest NMR results \cite{Chronister2020}.

Another possible avenue is worthy of consideration. The microscopic mechanism causing enhanced muon spin relaxation remains uncertain – scanned magnetic probe experiments do not resolve fields on the scale indicated by the muon spin relaxation – and so it in principle remains possible that the $\mu$SR results indicate some other transition in Sr$_2$RuO$_4$ that is not related to the superconductivity. If the $\mu$SR data are neglected, then there are also several further possibilities.  One is that transition splitting exists, but is much smaller under $[100]$ than $[110]$ uniaxial pressure, even as the van Hove singularity is approached. Notice this option is to leading order in strain not allowed for $B_{2g}$-nematic $E$-pairing  or  TRSB superconductivity in the $E$-representation.  Another possibility is a $B_{1g}$ nematic superconductor in the $E$-representation that yields a discontinuity of the elastic constants without split transitions or TRSB\ \cite{Benhabib2020} (see also \textit{SI Appendix, Heat Capacity Anomaly with $E_{g}$ or $E_{u}$ Pairing under Strain}). Such a state should be detectable via in-plane anisotropies below $T_c$, similar to the recently observed behavior in doped Bi$_2$Se$_3$ \cite{Matano2016,Yonezawa2017,Cho2020}.

A third possibility consistent with these heat capacity results considered in isolation is that the order parameter is single component at zero strain.  However, this would require re-interpretation of both the recent $\mu$SR and ultrasound results; see  Ref.\ \cite{Willa2020}  for a discussion.

\section*{Conclusion}

In summary, our comprehensive, high-resolution measurements of the heat capacity of Sr$_2$RuO$_4$ as a function of strain, temperature and magnetic field confirm that previous susceptibility measurements were probing a bulk rather than a surface phenomenon. We do not resolve any signs of transition splitting of two order parameter components. Analysis of the data in comparison with recent findings from other thermodynamic and spectroscopic measurements places strong constraints on the search for a full theoretical explanation of its order parameter. Our results further guide that search by highlighting the importance of order parameters that can be maximal at the location of the van Hove singularity, and point to areas in which further experimental work is desirable.

\matmethods{
High-quality single-crystal Sr$_2$RuO$_4$ samples were grown in a floating zone furnace (Canon Machinery) using techniques refined over many years to those described recently in \cite{Bobowski2019}.  They were aligned using a bespoke Laue x-ray camera, and cut using a wire saw into thin bars with whose long axis aligned with the $[100]$ direction of the crystal.  For the best results these bars were polished using home-made apparatus based on diamond impregnated paper with a minimum grit size of 1~$\mu$m. The bar was then mounted within the jaws of a uniaxial pressure rig using Stycast 2850FT epoxy (Henkel Loctide). A resistive thin film resistor chip (State of the Art, Inc.) as heater and a calibrated Au-AuFe($0.07\%$) thermocouple were fixed to opposite sides of the sample using Dupont 6838 silver epoxy. Special care was taken when epoxying to the pressure cell to minimize tilt and ensure as homogeneous a strain field as possible.
The uniaxial pressure apparatus was mounted on a dilution refrigerator, with thermal coupling to the mixing chamber via a high purity silver wire.  The data shown in the paper were acquired between 500~mK and 4.2~K, with operation above 1.5~K achieved by circulating a small fraction of the mixture.  The thermocouple was spot-welded in-house and its calibration fixed by reference to that of a calibrated RuO$_2$ thermometer.  The extremely low noise level of 20~pVHz$^{-1/2}$ on the thermocouple readout was achieved by the combination of a low temperature transformer (CMR direct) mounted on the 1~K pot of the dilution refrigerator, operating at a gain of 300, and an EG\&G 7265 lock-in amplifier. A Keithley 6221 low-noise current source was used to drive the heater.  The piezo-electric actuators were driven at up to $\pm 400$~V using a bespoke high-voltage amplifier.
In a setup of this kind, the significance of heat leaks to the environment is gauged by the lower cut-off frequency of set-up response curves.  By taking data at frequencies an order of magnitude higher than that lower cut-off, we ensure that the effect of such leaks makes a negligible contribution to our data. For further details on the technique see \cite{Li2020}.
}

\showmatmethods 

\acknow{We thank I.\ I.\ Mazin for useful discussions and R.\ Borth, M.\ Brando and U.\ Stockert for experimental support. Parts of this work were funded by the Deutsche Forschungsgemeinschaft (DFG, German Research Foundation) - TRR 288 - 422213477 (projects A10 and B01). NK acknowledges the support from JSPS KAKENHI (nos. JP17H06136 and JP18K04715) and JST-Mirai Program (no. JPMJMI18A3) in Japan and YM from JSPS KAKENHI (nos. JP15H05852, JP15K21717) and JSPS core-to-core programme.  YSL acknowledges the support of a St Leonard’s scholarship from the University of St Andrews, the Engineering and Physical Sciences Research Council via the Scottish Condensed Matter Centre for Doctoral Training under grant EP/G03673X/1, and the Max Planck Society.}

\showacknow 

\pnasbreak



\renewcommand{\figurename}{Fig.\ S\!\!}
\renewcommand{\theequation}{S\arabic{equation}}

\clearpage


\section*{Supporting Information Appendix}
\vspace{1em}

\setcounter{equation}{0}

\subsection*{Heat Capacity Measurements in the Temperature Range between 0.5 and 4 K}

\begin{figure}[b!]
\includegraphics[width=0.9\linewidth]{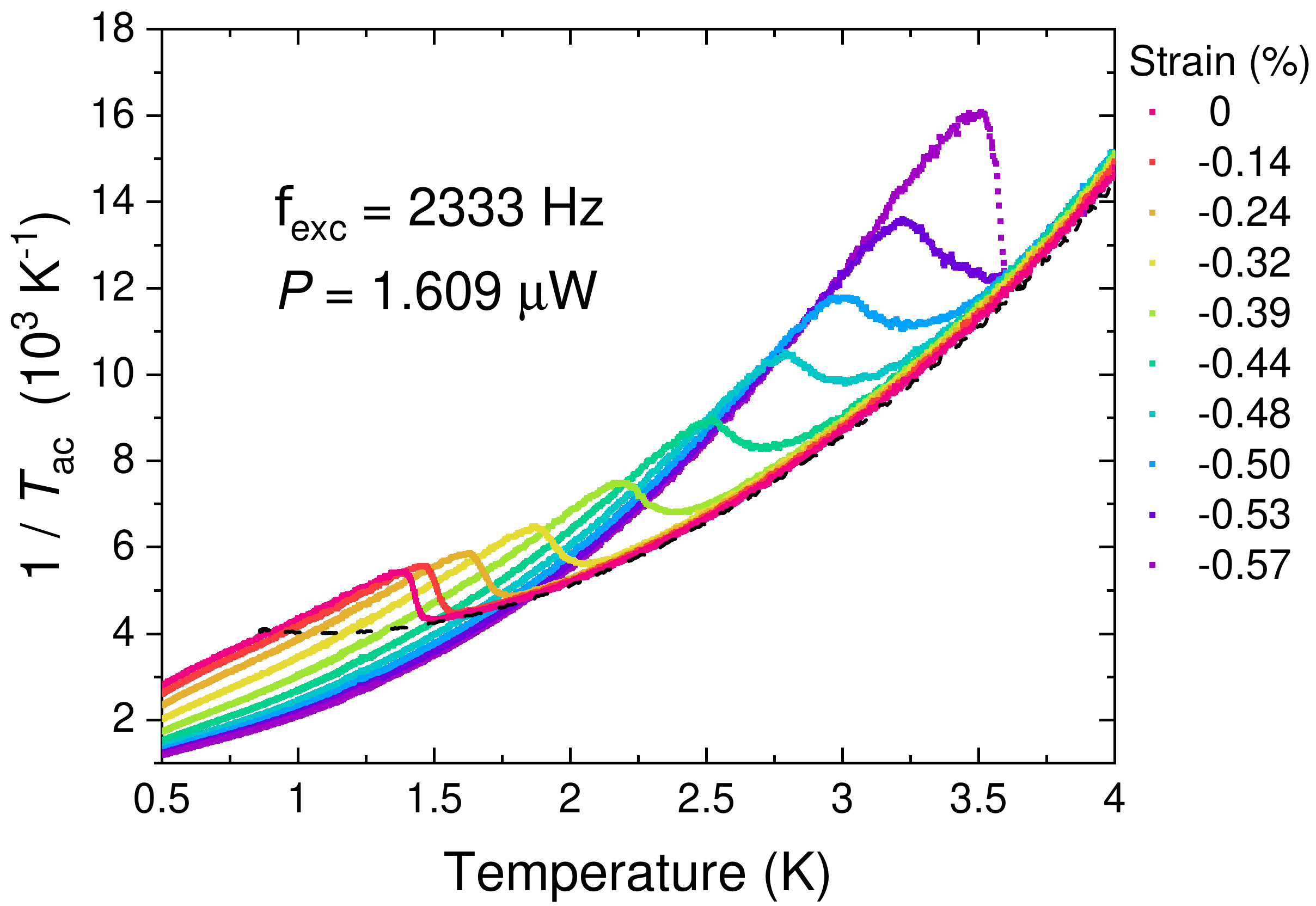}
\centering
\caption{
Heat capacity measurements for sample S4 at $f_{exc} = 2333$~Hz under various strains. $1/T_{ac}$ against temperature for different strains up to the peak in $T_c$. The dashed line is a heat capacity measurement at $\mu_0H_{\parallel c} = 0.1$~T and $\varepsilon_{xx} = 0\%$.
}
\label{HC_full}
\end{figure}

In the main text, we have shown the heat capacity measurements between 1 and 4~K taken at a frequency of 3913~Hz. In Fig.~S\ref{HC_full} we present additional data at 2333 Hz in an extended temperature range down to 0.5~K. For each strain two data sets recorded in different temperature regions, from 0.5 to 1.3~K and from 1 to 4~K, have been combined. The data sets of the two different measurement runs coincide in an excellent way in the overlap region. The increase in $T_c$ with strain is consistent with the result at $f_{exc} = 3913$~Hz. Within our experimental resolution, we do not find any indication of an additional anomaly below the superconducting transition at any given strain.

\subsection*{Heat Capacity Jump}

\begin{figure}[b!]
\includegraphics[width=0.9\linewidth]{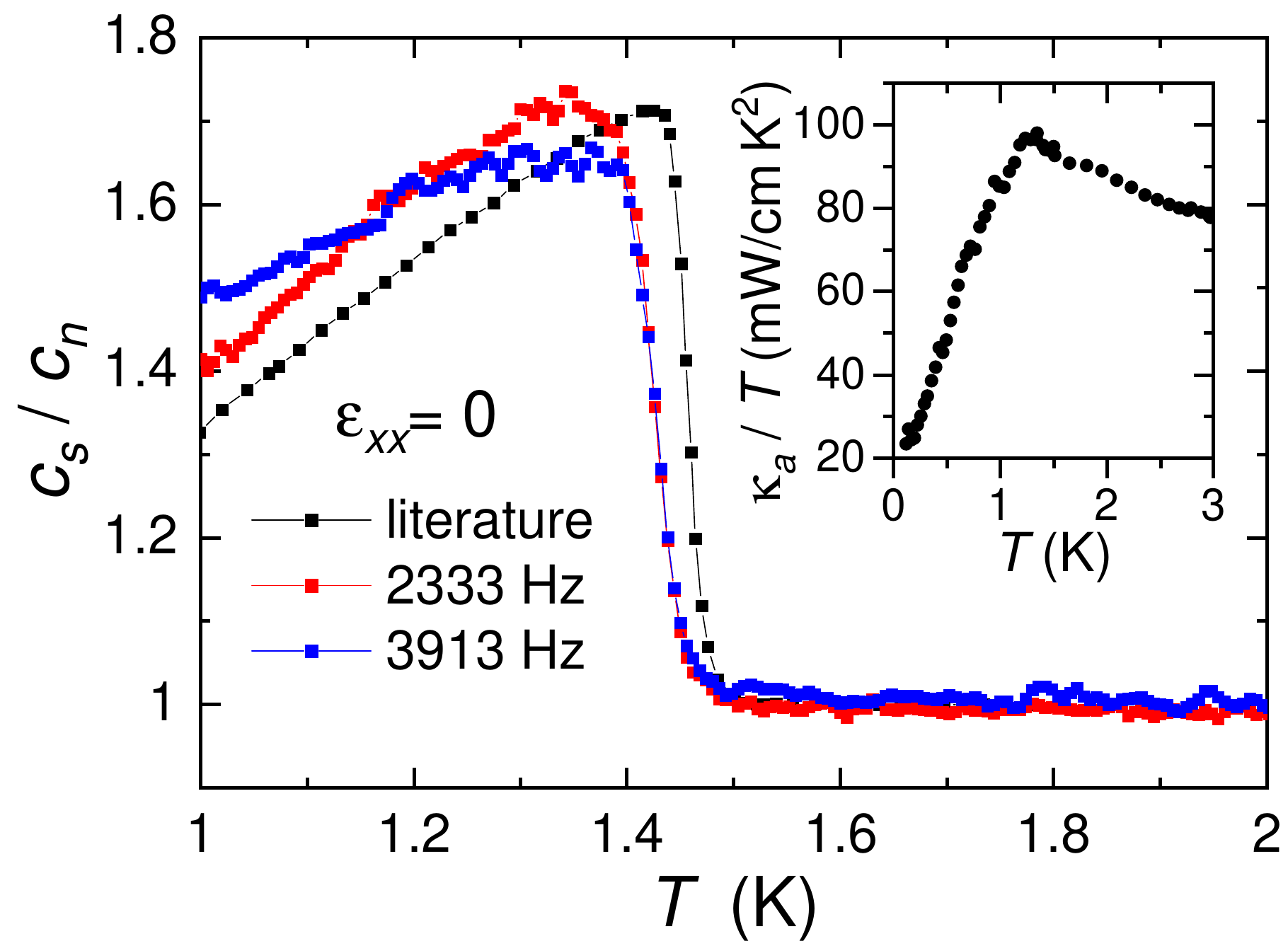}
\centering
\caption{
Calculations of normalized specific heat curves at zero strain from our ac calorimetry by using Eq.\ \ref{S3} and the thermal conductivity from \cite{Hassinger2017} reproduced in the inset. The red and blue curves are the calculations for S4 with $f_{exc} = 2333$~Hz and 3913~Hz, respectively. The black curve is the published data from Deguchi \textit{et al.}\ \cite{Deguchi2004}.
}
\label{HC_jump}
\end{figure}

Under an assumption that the heater is narrow and the cross-sectional area $A$ of the sample is small in comparison with the thermal diffusion length as in our case, the measured signal $T_{AC}(T)$ is related to the volume specific heat capacity $c_v$ by
\begin{equation}\label{S3}
  c_{v}(T)=\left (\frac{P \times F(\omega)}{2A}\right )^2 \times \frac{1}{\omega\times 2\kappa(T)} \times \frac{1}{[T_{AC}(T)]^2}.
\end{equation}	

Although we do not have sufficient information to calculate $c_v(T)$ at all strains, enough is known about the thermal conductivity $\kappa(T)$ at zero strain to allow this to be attempted (see inset of Fig.\ S\ref{HC_jump} and \cite{Hassinger2017}).  The results, shown in Fig.\ S\ref{HC_jump}, demonstrate the level of accuracy that can be achieved using the present set-up for which Eq.\ \ref{S3} is a reasonable description.
These results prove that although the full deduction of the volume specific heat capacity is not necessary for the physics that we sought to investigate in the current experiments, it will be possible in principle to determine it with good accuracy.

In the case of a second-order superconducting phase transition $\kappa(T)$ is continuous across the transition and we can determine the evolution of the anomaly height $\Delta c_s /c_n$ with $T_c$ from our data despite not knowing the volume specific heat. Using Eq.~\ref{S3} and the fact that $\kappa_n\approx\kappa_s$. We find for the superconducting state heat capacity normalized to the normal-state value:
\begin{equation}\label{S4}
  \frac{c_{v}^s}{c_{v}^n}=\frac{\kappa_n}{\kappa_s} \times \left(\frac{T_{AC}^n}{T_{AC}^s}\right)^2 \approx \left(\frac{T_{AC}^n}{T_{AC}^s}\right)^2
\end{equation}	
The indices $s$ and $n$ indicate the corresponding values in the superconducting and in the normal state, respectively. More details can be found in \cite{Li2020}.

\subsection*{Experimental Limits}

\subsubsection*{Experimental limits on the detection of a potential second superconducting transition}

\begin{figure}[b!]
\includegraphics[width=0.95\linewidth]{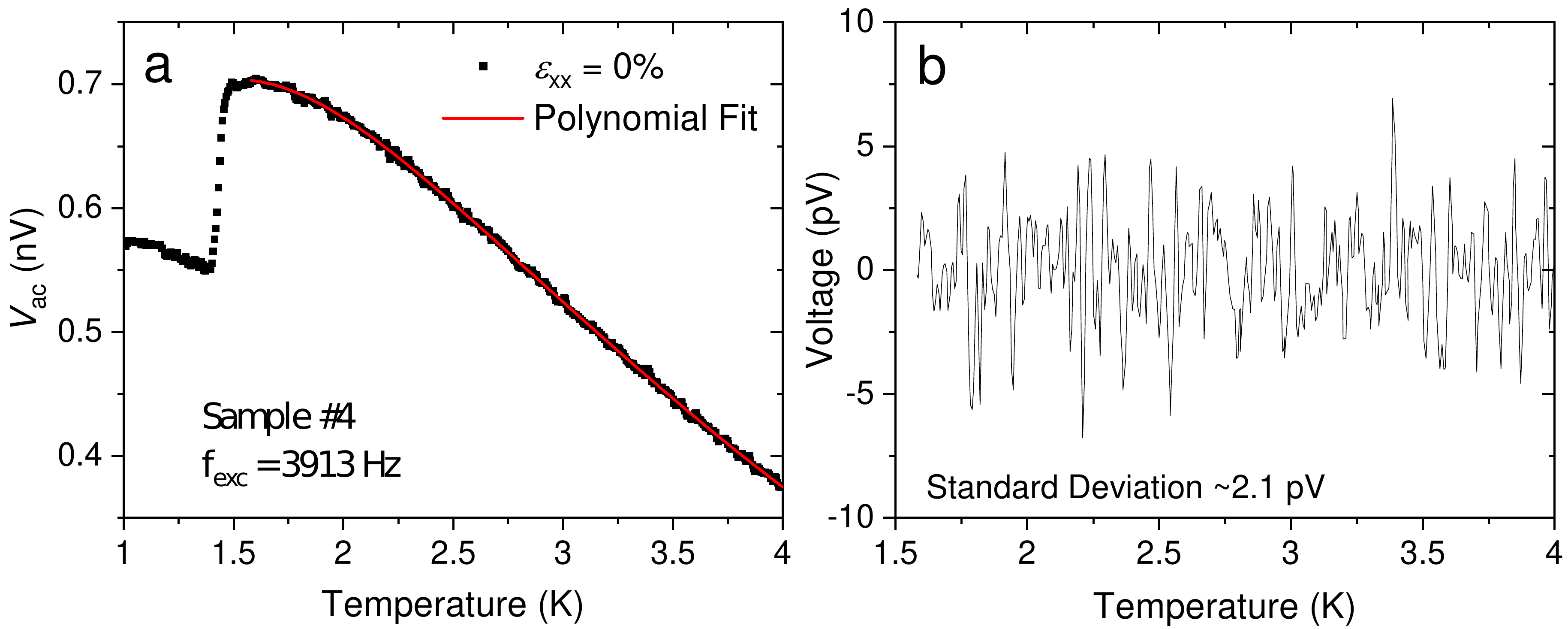}
\centering
\caption{
Raw signal and noise. (a) The thermocouple voltage readout $V_{ac}$ against temperature at zero strain for S4. The red line is a fit to the normal state data with a $5^{\rm th}$ degree polynomial. (b) The difference between the measured signal $V_{ac}$ and the fitted curve. The standard deviation is about 2.1~pV.
}
\label{Raw}
\end{figure}

The signal-to-noise ratio determines one of the experimental limits on detecting a potential second superconducting transition in Sr$_2$RuO$_4$. The noise level remains almost the same for all strains; therefore, we used the heat capacity data at zero strain to analyze the experimental resolution limit given by the noise of the thermometer readout. Fig.~S\ref{Raw}a shows the raw signal $V_{ac}$ of the thermocouple as function of temperature. The high sensitivity voltage readout was achieved by a transformer ($300\times$ amplification) mounted on the 1~K pot of the dilution refrigerator and using time constants of 20 and 50~s for 2333 and 3913~Hz, respectively. To determine the noise we fit a polynomial to the normal state data and subtracted it. Fig.~S\ref{Raw}b shows the result as function of temperature. The noise is temperature-independent and its standard deviation $\sigma$ is only 2.1~pV at the thermocouple.

\begin{figure}[t!]
\includegraphics[width=0.9\linewidth]{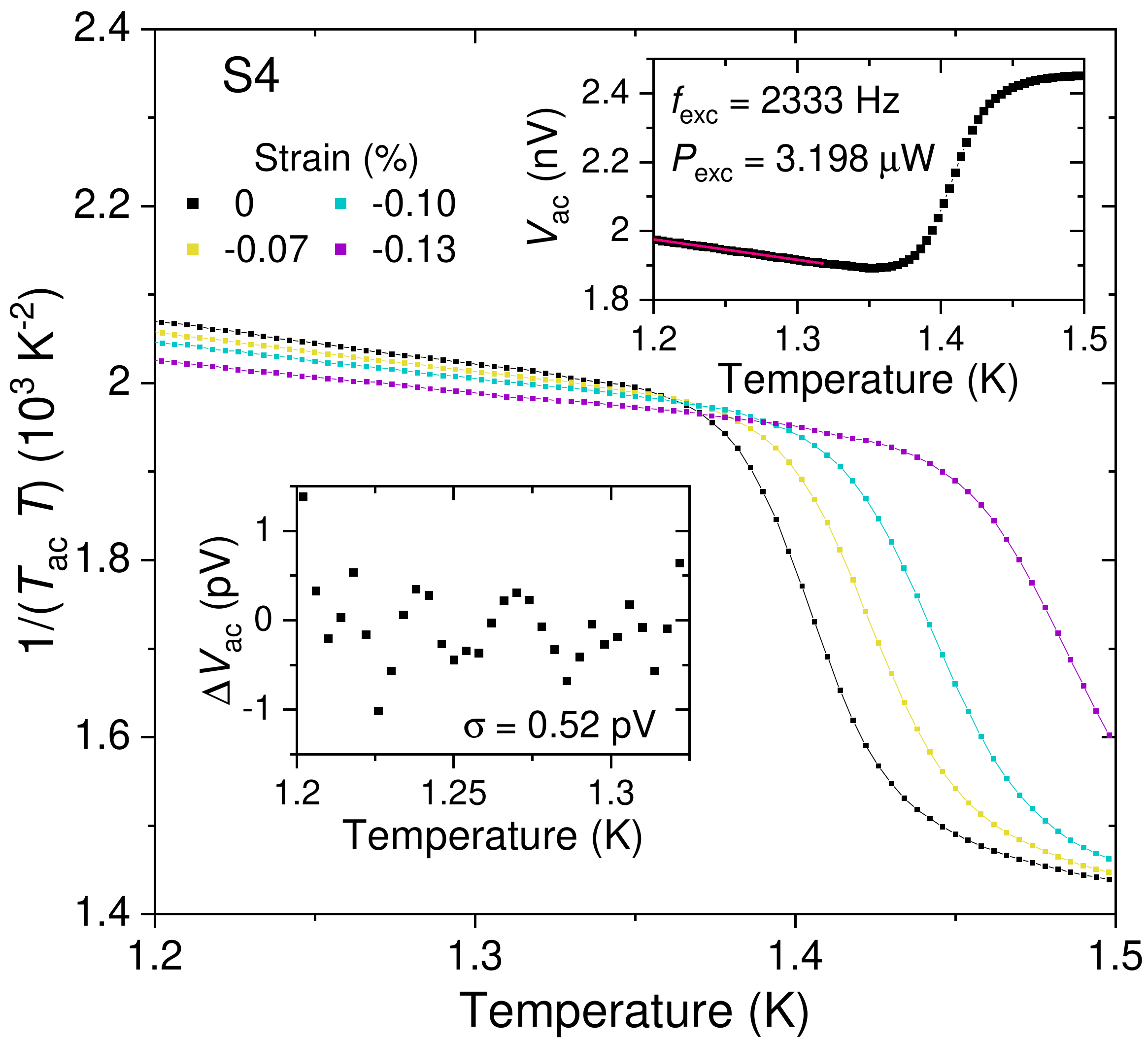}
\centering
\caption{
High resolution heat capacity measurements on sample S4. Measurements at several low strains. $P_{exc}$ was doubled and the measurements were repeated several times at each strain 20, 34, 18 and 20 times for the 0\%, -0.07\%, -0.1\% and -0.13\% curves, respectively. (upper inset) Averaged voltage readout from the thermocouple at zero strain. The red line is a linear fit to the data below $T_c$. (lower inset) The difference between the measured curve and the linear fit. The standard deviation is only about 0.52~pV.
}
\label{HC_precision}
\end{figure}

Since $95\%$ of data points fall within twice the standard deviation in a normal distribution, we use $4\sigma\approx8.4$~pV as the detection limit for a step size. The size of the signal at $T = 1.6$~K is about 0.7~nV so the experimental limit on detecting a small jump in $\Delta c/c$ is $\delta V_{ac}/V_{ac}=8.4~{\rm pV} /0.7~{\rm nV}=0.012$. The visible transition in the data has a jump size of $\Delta c/c \approx0.3$. That implies that a potential second transition has to be more than $0.012/0.3=4\%$ of the size of the visible transition to be resolved. Therefore, the experimental limit on detecting the second transition is about $4\%$ of the visible one for S4 with $f_{exc} = 3913$~Hz in the temperature region around 1.6~K.

In Fig.\ S\ref{HC_precision} we present an analysis using a set of heat capacity recorded with a doubled excitation power. Additionally, the measurement runs were repeated and averaged up to 34 times providing higher quality heat capacity data with a much better signal-to-noise ratio. We restricted the measurements to the temperature range between 1.2 and 1.5 K since a potential second transition is expected at temperatures lower than the visible transition anomaly. There is also no signature of a second transition in the averaged curves. The noise of the averaged voltage at the thermocouple is only 0.52~pV (see lower inset of Fig.\ S\ref{HC_precision}). Following the analysis scheme described above, the detection limit relative to the primary transition is only 0.3\%. We note, however, that this value concerns the ability to resolve a sharp discontinuity.

\begin{figure}[tb!]
\includegraphics[width=0.85\linewidth]{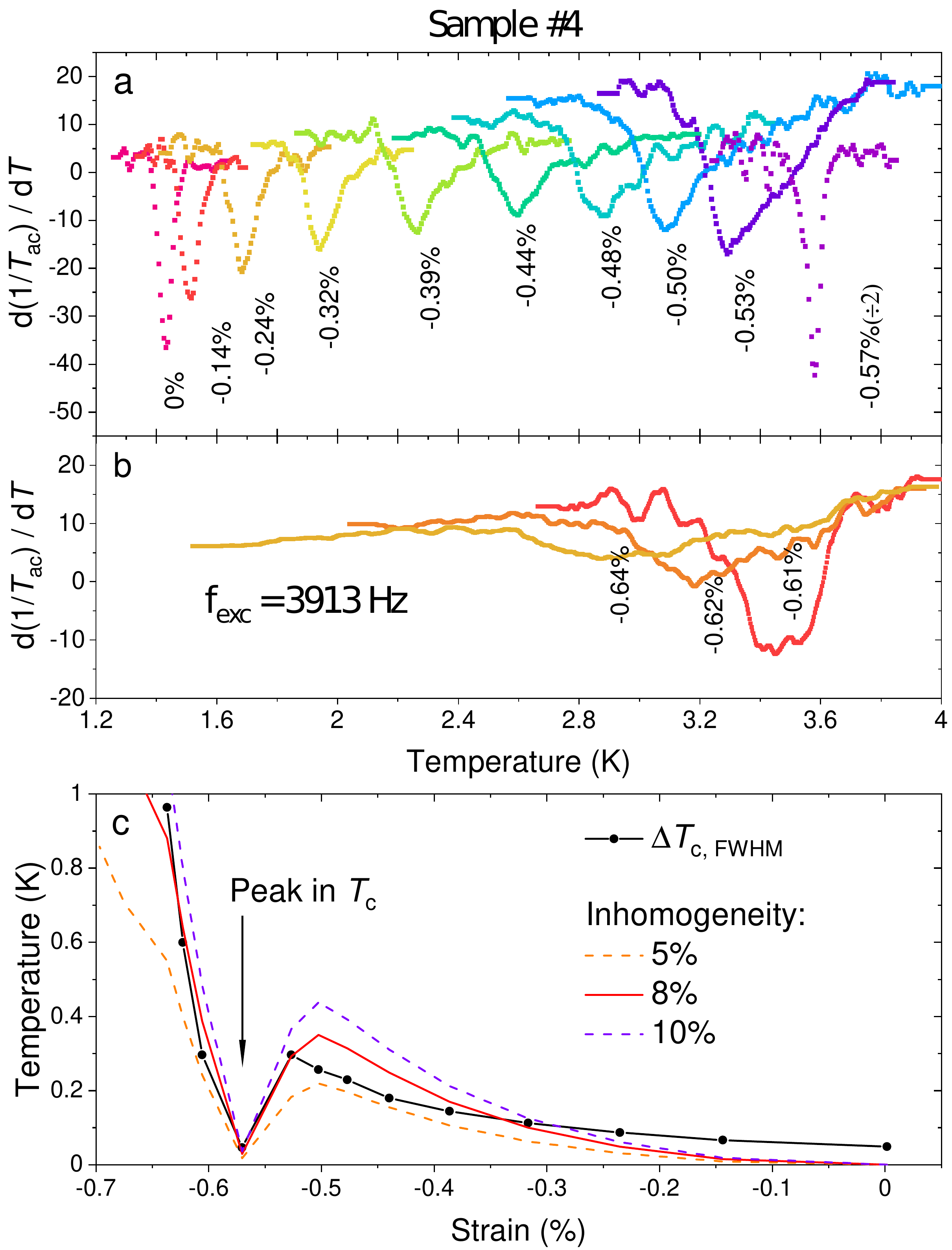}
\centering
\caption{
Temperature resolution and strain inhomogeneity. The first derivative of $1/T_{ac}$ with respect to temperature for S4 at different strains (a), before the peak in $T_c$ and (b), after the peak. The curve at $-0.57\%$ in panel a is reduced by a factor of 2 for clarity. (c) The transition breadth against strain. The solid points are the FWHM derived from the results in (a) and (b). Three simulation curves (see text) with different strain inhomogeneities are shown for comparison. The arrow marks the position of the peak in $T_c$.
}
\label{ResStrain}
\end{figure}

\subsubsection*{The limit on determining the separation of two potential transitions}
It will be not possible to resolve two transitions if they are too close together. Therefore, in Sr$_2$RuO$_4$ the breadth of the visible transition determines the detection limit for the separation of two transitions. Figures~S\ref{ResStrain}a and S\ref{ResStrain}b show the first derivative of the measured $1/T_{ac}(T)$ data with $f_{exc} = 3913$~Hz at strains before and after the peak in $T_c$ for S4, respectively. The full width at half maximum (FWHM) is used as a conservative criterion for the limit on determining the separation of two potential transitions. The results are shown in Fig.~S\ref{ResStrain}c. The FWHM increases from 50~mK at zero strain to 300~mK at $\varepsilon_{xx} = -0.53\%$.

\subsubsection*{Strain inhomogeneity}
Strain inhomogeneity causes a distribution of $T_c$'s and leads to a rounded heat capacity anomaly. From the breadth of the transition, the size of the inhomogeneity can be estimated as follows:
\begin{equation}\label{dTc}
\Delta T_c\cong \frac{dT_c(\varepsilon)}{d\varepsilon}\times \Delta \varepsilon =\frac{dT_c(\varepsilon)}{d\varepsilon}\times \frac{\Delta \varepsilon}{\varepsilon} \times\varepsilon.
\end{equation}	
$\Delta T_{c,{\rm FWHM}}$ can be determined from the $d(1/T_{ac})/dT$ curves as shown in Fig.~S\ref{ResStrain} and, therefore,
\begin{equation}\label{dTcFWHM}
\Delta T_{c,{\rm FWHM}}=\left|\frac{dT_c(\varepsilon)}{d\varepsilon}\right| \times \frac{\Delta \varepsilon_{FWHM}}{|\varepsilon|}\times |\varepsilon|.
\end{equation}
The distribution in $T_c$, $\Delta T_{c,{\rm FWHM}}$, is related to the tangent slope on the $T_c(\varepsilon)$ curve $|dT_c(\varepsilon)/d\varepsilon|$, the strain inhomogeneity $\Delta\varepsilon_{FWHM}/|\varepsilon|$  and the applied strain $|\varepsilon|$. It is inevitable to have a certain strain inhomogeneity in a sample and, therefore, the higher the applied strain, the wider the distribution in $T_c$. Note, the distribution is even larger when the applied strain goes beyond $\varepsilon_{{\rm peak\,in}\,T_c}$  because the tangent slope is steeper. The $T_c(\varepsilon)$ curve determined by the midpoints of the leading edge of the transitions is used to simulate $\Delta T_{c,{\rm FWHM}}$ with different sizes of the inhomogeneity as shown in Fig.~S\ref{ResStrain}c. The values of $\Delta T_{c,{\rm FWHM}}$ near zero strain are different from the simulations because there is a finite transition width intrinsic to the sample. The distribution of $T_c$ scales with the tangent slope on $T_c(\varepsilon)$. Hence, the inhomogeneity determined primarily by matching to the simulations around the peak in $T_c$ is approximately $8\%$ in sample S4.

\subsection*{Heat Capacity Anomaly with {\boldmath $E_{g}$ or $E_{u}$} Pairing under Strain}

We summarize the heat capacity anomalies under strain
for a system that at zero strain is in an $E_{g}$ or $E_{u}$ pairing
state of the point group $D_{4d}$. We will analyze all three options
for such pairing state, the chiral time-reversal symmetry breaking
states or the nematic states that are all doubly degenerate at zero
strain:
\begin{eqnarray}
\left(\psi_{x},\psi_{y}\right) & \propto & \left(1,i\right)\,{\rm or}\,\,\left(1,-i\right)\,\,{\rm TRSB\,\,superconductor} \\
& \propto & \left(1,1\right)\,{\rm or}\,\left(1,-1\right)\,B_{2g}{\rm -nematic\,\,superconductor}\nonumber \\
& \propto & \left(1,0\right)\,{\rm or}\,\left(0,1\right)\,\,\,\,\,B_{1g}{\rm -nematic\,\,superconductor}.\nonumber
\end{eqnarray}
For all three states the degeneracy is lifted at finite  $B_{1g}$ or  $B_{2g}$ strain. We
will focus in particular on strain that transforms according to $B_{1g}$
symmetry.

For the chiral $\left(1,\pm i\right)$ and nematic $\left(1,\pm1\right)$
states we will see that the slopes of the phase boundaries are tied
to the heat capacity anomalies at those phase transitions. This allows
us to check whether the observed phase boundaries of the strain split
transition in Sr$_{2}$RuO$_{4}$ are consistent with the observed heat capacity
data. Our conclusion is that no consistent description seems to exist,
at least if one is limited to the leading-order strain expansion.
If this expansion is applicable, existing observations rule out that
the order parameter of Sr$_{2}$RuO$_{4}$ transforms according to a two-dimensional
irreducible representation with chiral $\left(1,\pm i\right)$ and
nematic $\left(1,\pm1\right)$ order. We also discuss the behavior
for the $\left(1,0\right)$, $\left(0,1\right)$ transition where
finite strain breaks the underlying nematic state.

\begin{figure}[b!]
\includegraphics[scale=0.6]{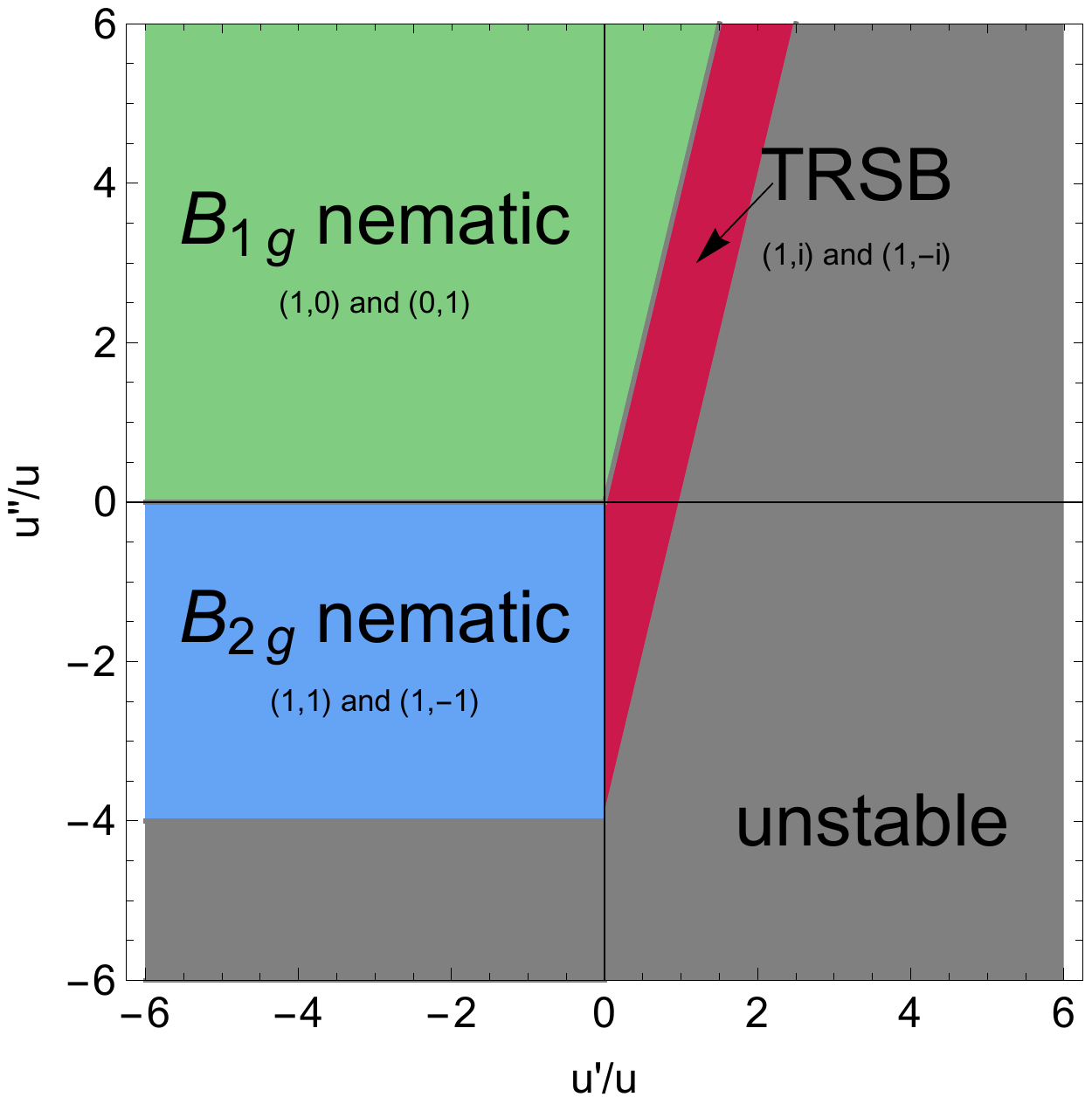}

\caption{Phase diagram of the Ginzburg Landau expansion Eq.~\ref{eq:GL} for the different pairing states in the $E$-representation as
function of the ratios $u'/u$ as well as $u''/u$ of the quartic
coupling constants. Depending on the values of the quartic interactions,
three possible superconducting states are allowed. In the region marked
``unstable'' the global minimum is at infinity and higher order
terms need be included to describe a first order transition.}

\label{Fig_phase_diagram}
\end{figure}

The Ginzburg-Landau expansion of the $E_{g}$ or $E_{u}$ pairing
states is
\begin{eqnarray}
F & = & F_{0}+\frac{a\left(T\right)}{2}\left(\left|\psi_{x}\right|^{2}+\left|\psi_{y}\right|^{2}\right)+\frac{u}{4}\left(\left|\psi_{x}\right|^{2}+\left|\psi_{y}\right|^{2}\right)^{2} \nonumber \\
 & + & \frac{u'}{4}\left(\psi_{x}^{*}\psi_{y}-\psi_{y}^{*}\psi_{x}\right)^{2}+\frac{u''}{4}\left|\psi_{x}\right|^{2}\left|\psi_{y}\right|^{2}.\label{eq:GL}
\end{eqnarray}
In the case of $E_{g}$-pairing, $\psi_{x}$ and $\psi_{y}$ correspond
to the amplitudes of the $d_{xz}$ or $d_{yz}$ components of the
order parameter, respectively. For $E_{u}$ pairing they correspond
to the weights of $p_{x}$ and $p_{y}$ pairing. As usual, we use
$a\left(T\right)=a_{0}\left(T-T_{c,0}\right)$.

To ensure that the Ginzburg Landau expansion is globally stable it
must hold that $u>0$. In addition, for $u'>0$ it is necessary that
$u-u'+\frac{1}{4}u''>0$, while for $u'<0$ it must hold that $u'>-4u$.
When these conditions are not fulfilled the transition is of first
order, which is the region labelled {\em unstable} in Fig.~S\ref{Fig_phase_diagram}. To obtain these conditions we write $\psi=\psi_{0}\left(\cos\theta,\sin\theta e^{i\varphi}\right)$
and obtain for the quartic term
\begin{equation}
F_{4}=\frac{\psi_{0}^{4}}{4}\left[u+\left(\frac{u'}{2}\cos2\varphi-\left(\frac{u'}{2}-\frac{u''}{4}\right)\sin^{2}\left(2\theta\right)\right)\right].
\end{equation}
Under the above conditions this expression is guaranteed to be positive
definite for all $\theta$ and $\varphi$. This expansion also demonstrates
that sign of $u'$ is important to obtain a phase $\varphi=n\pi$,
which preserves TRS or $\varphi=\left(n+\frac{1}{2}\right)\pi$ which
breaks TRS. The TRSB chiral state $\left(1,\pm i\right)$ has lowest
energy for $u'>0$ and $4u'>u''$. The $\left(1,\pm1\right)$ nematic
has lowest energy for $u'<0$ and $u''<0$. Finally, the $\left(1,0\right)$,
$\left(0,1\right)$ nematic is the most stable solution for $4u'<u''$
and $u''>0$. The various regions of stability are shown in Fig.\ S\ref{Fig_phase_diagram}.

When we add the strain $\epsilon_{B_{1g}}\equiv\frac{1}{2}\left(\epsilon_{xx}-\epsilon_{yy}\right)$
of $B_{1g}$ symmetry a linear in $\epsilon_{B_{1g}}$ term
\begin{equation}
f_{\epsilon}=-\frac{\lambda}{2}\epsilon_{B_{1g}}\left(\left|\psi_{1}\right|^{2}-\left|\psi_{2}\right|^{2}\right)
\end{equation}
is allowed. It lifts the degeneracy of the two-component representation.

\begin{figure*}[tb!]
\includegraphics[scale=0.45]{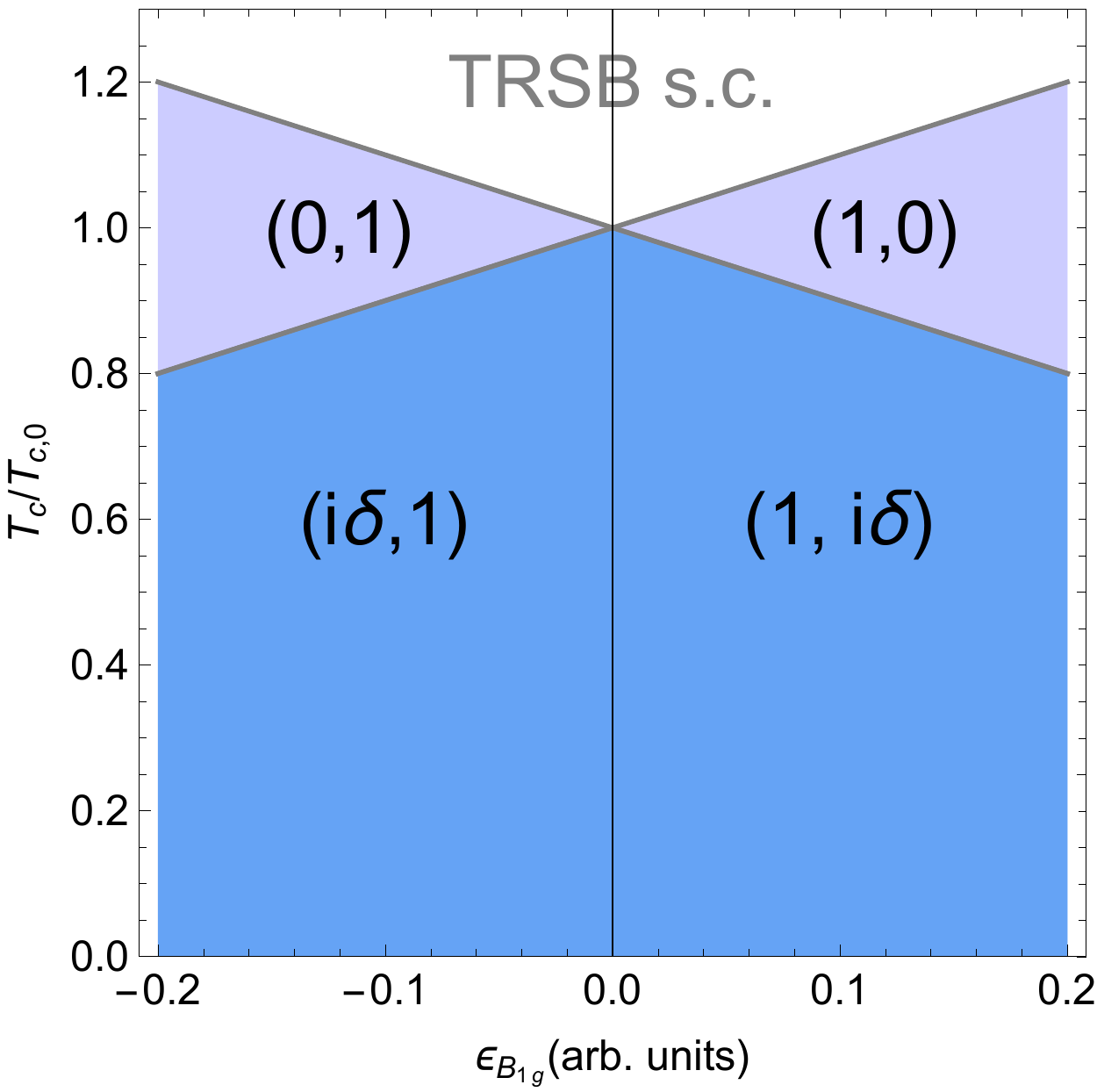}\includegraphics[scale=0.45]{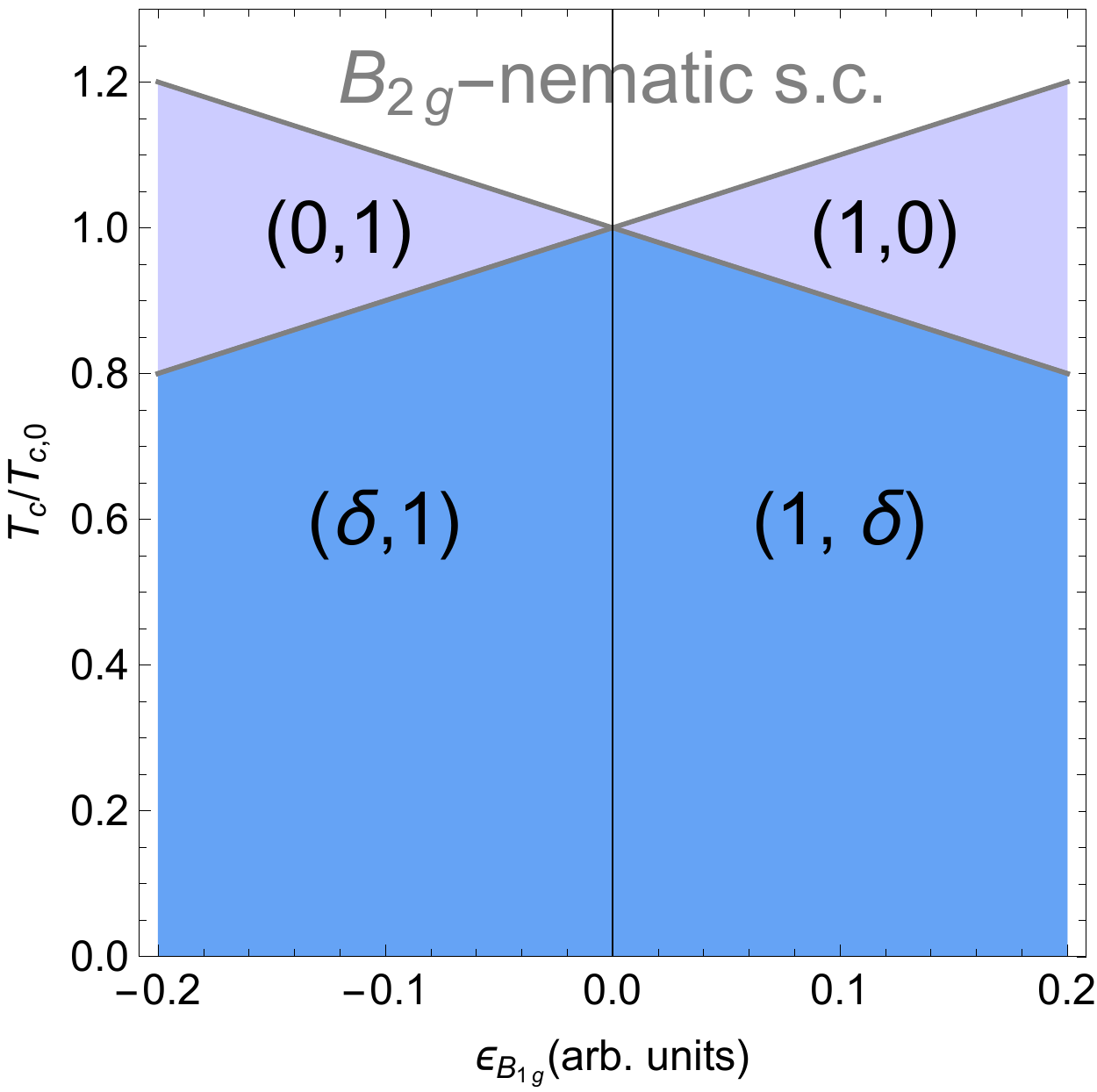}\includegraphics[scale=0.45]{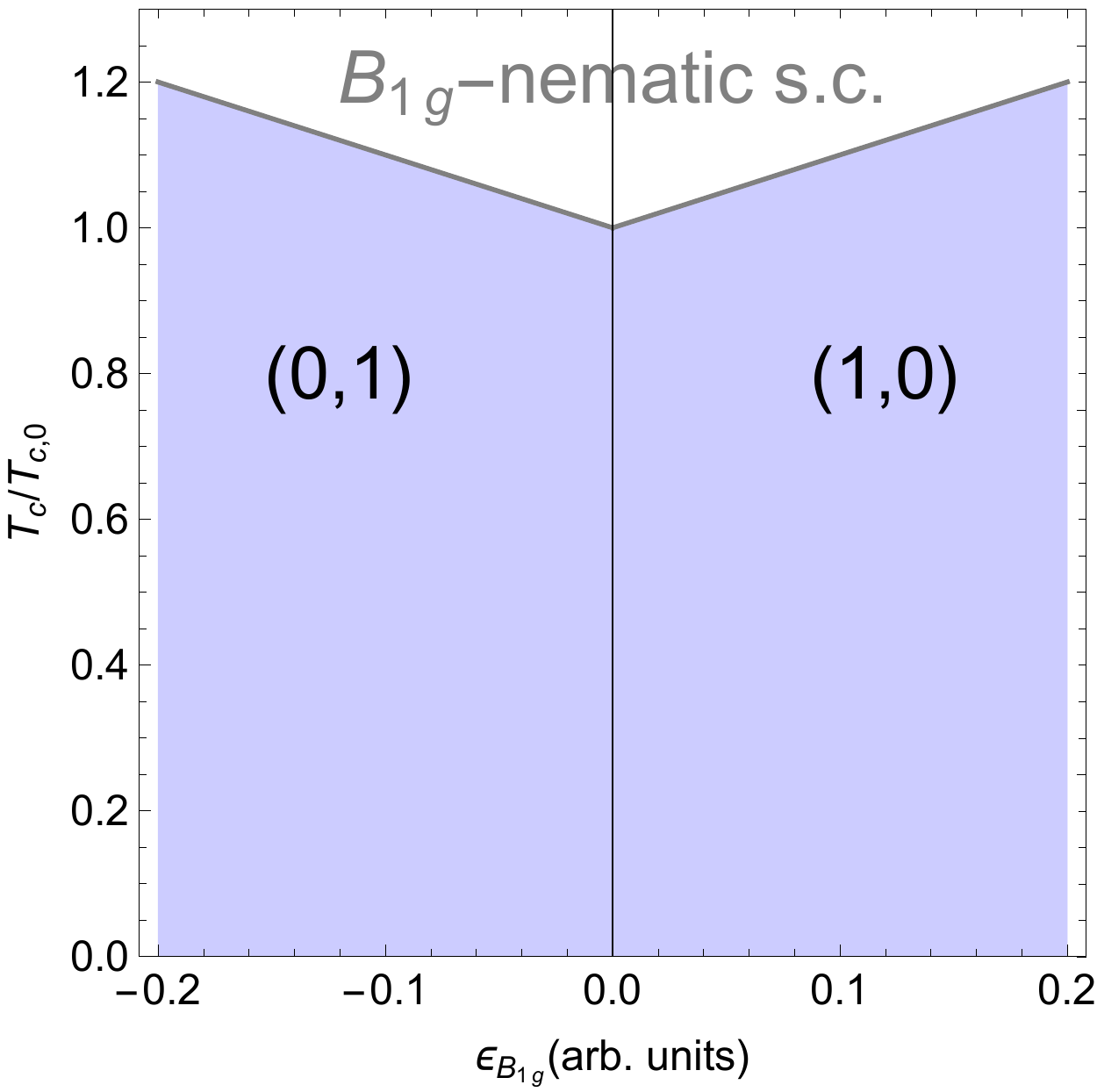}

\caption{Temperature $B_{1g}$ strain phase diagram for the three possible
pairing states with $E_{g}$ or $E_{u}$ pairing symmetry. At $\epsilon_{B_{1g}}=0$
all three states are doubly degenerate. For the TRSB superconductor
the degenerate states are $\left(1,\pm i\right)$ for the $B_{2g}$
nematic superconductor they are $\left(1,\pm1\right),$ while for
the $B_{1g}$ nematic superconductor the degenerate states are $\left(1,0\right)$
and $\left(0,1\right)$. At finite strain the TRSB and $B_{2g}$ nematic
superconductors display split transitions. Superconductivity and the
corresponding global $U\left(1\right)$ symmetry breaking sets in
at the upper transition, while at the lower transition the system
breaks time reversal symmetry or rotational symmetry, respectively.
The additional component is then characterized by a relative strength
$\delta$ that vanishes continuously at the lower transition. It holds
$-1<\delta<1$ at finite strain and $\delta=\pm1$ at zero strain.
The situation is different for a $B_{1g}$ nematic superconductor.
At zero strain it spontaneously breaks rotation symmetry, i.e.\ it
is a nematic superconductor. This should be visible in thermal expansion
measurements, where the length changes below $T_{c}$ are differemt
along the $x$- and $y$-direction. Since non-zero external strain
explicitly breaks this symmetry, no second phase transition takes
place at a lower transition. Hence, in this case only one heat capacity
anomaly should be observable. However, applying $B_{2g}$ strain should
give rise to split transitions with two heat capacity anomalies.
For all three states two heat capacity anomalies in $C_{66}$ and $C_{11}-C_{12}$ should be visible.}
\label{Fig0}
\end{figure*}

In what follows we analyze all these states. The resulting phase diagrams
as function of strain are shown in Fig.\ S\ref{Fig0}

\subsubsection*{TRS-breaking chiral state}

If $u'>0$ and $u'>u''/4$, it follows that the ordered state breaks time
reversal symmetry
\begin{equation}
\left(\psi_{1},\psi_{2}\right)\propto\left(1,\pm i\right).
\end{equation}
Below the transition temperature $T_{c,0}$ the order parameter is
then given as
\begin{eqnarray}
\psi_{1} & = & e^{i\varphi}\sqrt{\frac{a_{0}}{u+w}}\left(T_{c,0}-T\right)^{1/2},\nonumber \\
\psi_{2} & = & \pm i \psi_{1} ,
\end{eqnarray}
where we introduced $w=u-2u'+\frac{1}{2}u''$. Hence, the conditions
$u'>u''/4$ and $u-u'+\frac{1}{4}u''>0$ translate to $u\pm w>0$.
The energy of this state is $F=F_{0}-\frac{a^{2}}{2\left(u+w\right)}$.

Next we consider the case with additional strain $\epsilon_{B_{1g}}$. Let us assume without restriction $\lambda\epsilon_{B_{1g}}>0$.
Above the upper transition
\begin{equation}
T_{c,u}=T_{c,0}+\frac{\lambda}{a_{0}}\epsilon_{B_{1g}}
\end{equation}
it holds $\psi_{1}=\psi_{2}=0.$ In the intermediate temperature regime
$T_{c,l}<T<T_{c,u}$ holds
\begin{eqnarray}
\psi_{1} & = & e^{i\varphi}\sqrt{\frac{a_{0}}{u}}\left(T_{c,u}-T\right)^{1/2}\nonumber \\
\psi_{2} & = & 0
\end{eqnarray}
where the lower transition temperature
\begin{equation}
T_{c,l}=T_{c,0}-\eta\frac{\lambda}{a_{0}}\epsilon_{B_{1g}}
\end{equation}
is determined by the dimensionless ratio of the interaction parameters
\begin{equation}
\eta=\frac{u+w}{u-w}>0.
\end{equation}
Notice, $\eta>1$ if $w>1$ and $\eta<1$ for $w<0$. A weakly strain
dependent lower transition temperature corresponds to $w\approx-u$.
Finally, below $T_{c,l}$ we have
\begin{eqnarray}
\psi_{1} & = & e^{i\varphi}\sqrt{\frac{a_{0}}{u+w}}\left(T_{c,l}+2\eta \frac{\lambda\epsilon_{B_{1g}}}{a_{0}}-T\right)^{1/2}\nonumber \\
\psi_{2} & = & \pm ie^{i\varphi}\sqrt{\frac{a_{0}}{u+w}}\left(T_{c,l}-T\right)^{1/2}.
\end{eqnarray}
In particular this yields $\psi_{1}\left(T_{x,l}\right)=e^{i\varphi}\sqrt{\frac{2\epsilon_{B_{1g}}}{u-w}}$.
The corresponding phase boundaries as function of strain are shown
in Fig.\ S\ref{Fig_phase_diagram}.

Inserting these results for the order parameters into the free energy
yields
\begin{equation}
F=F_{0}+\left\{ \begin{array}{ccc}
0 & {\rm if} & T_{c,u}<T\\
-\frac{\left(a-\lambda\epsilon_{B_{1g}} \right)^{2}}{4u} & {\rm if} & T_{c,l}<T<T_{c,u}\\
-\frac{a^{2}}{2\left(u+w\right)}-\frac{\lambda^{2}\epsilon_{B_{1g}}^{2}}{2\left(u-w\right)} & {\rm if} & T<T_{c,l}
\end{array}\right.
\end{equation}

\begin{figure*}[htb!]
\includegraphics[scale=0.45]{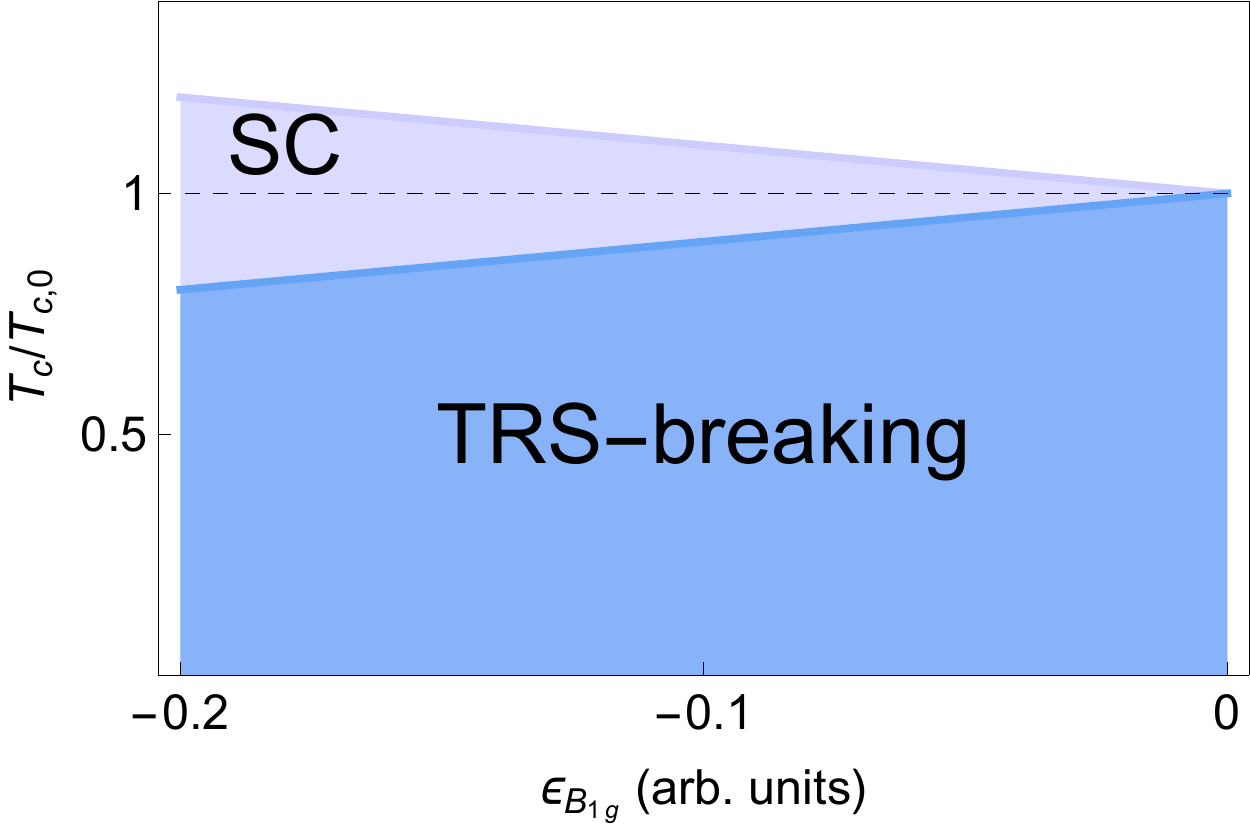}\includegraphics[scale=0.45]{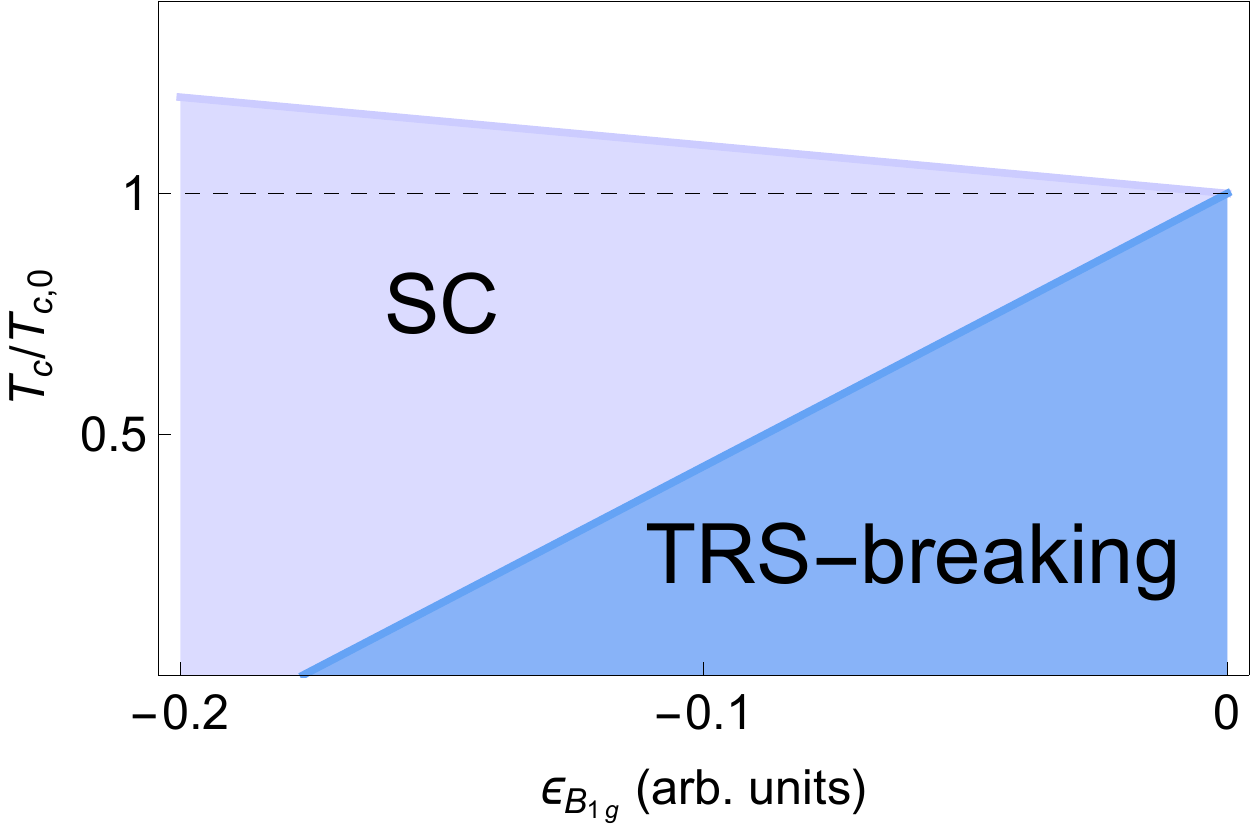}\includegraphics[scale=0.45]{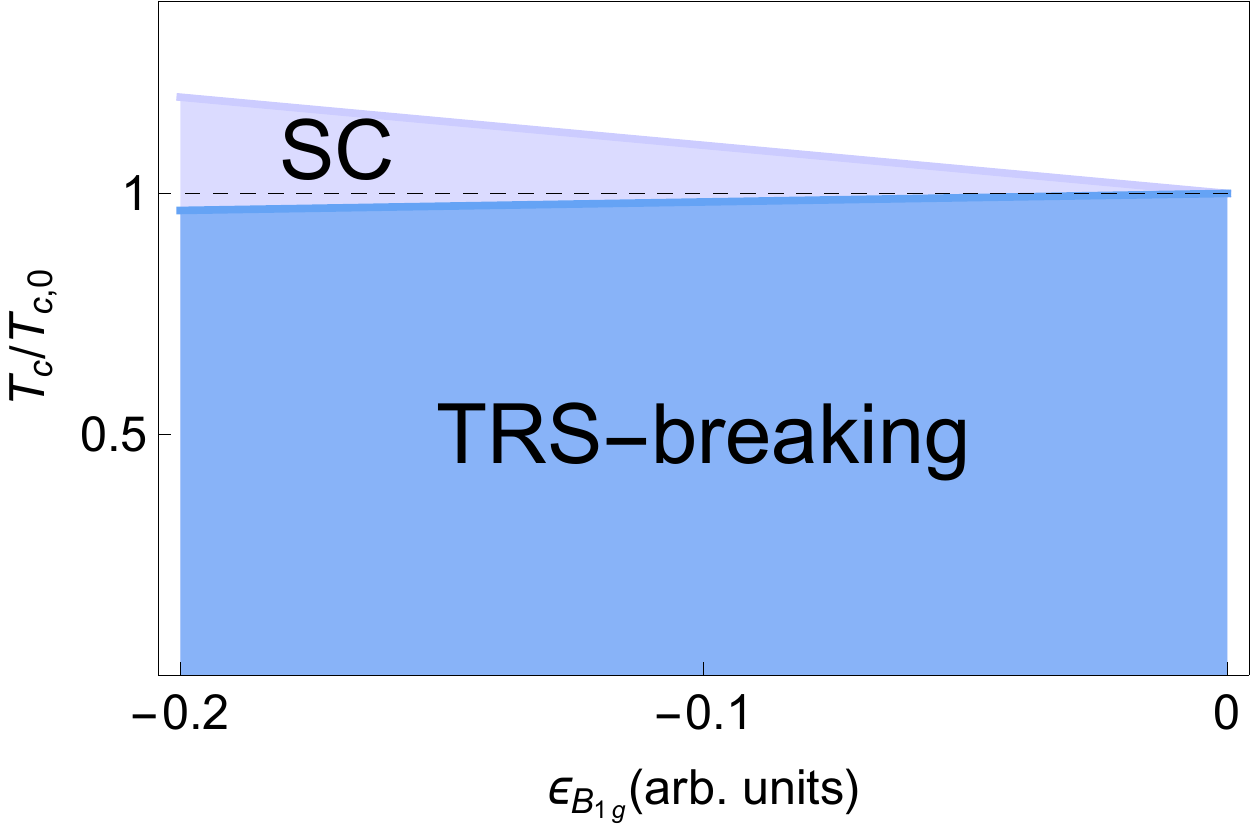}
\caption{The asymmetry in the strain dependence of the two transition temperatures
depends on the ratio of the interaction parameters $w/u$ of the Ginzburg-Landau
expansion. Left panel: $w=0$ with a symmetric splitting of both transition
temperatures, Middle panel: $w=0.7u$ where the lower transition temperatures
decreases more rapidly and Right panel: $w=-0.7u$ where the lower
transition temperature decreases only slightly. The latter seems to
be what is happening in Sr$_{2}$RuO$_{4}$.}\label{Fig1}
\end{figure*}

\begin{figure*}[htb!]
\includegraphics[scale=0.45]{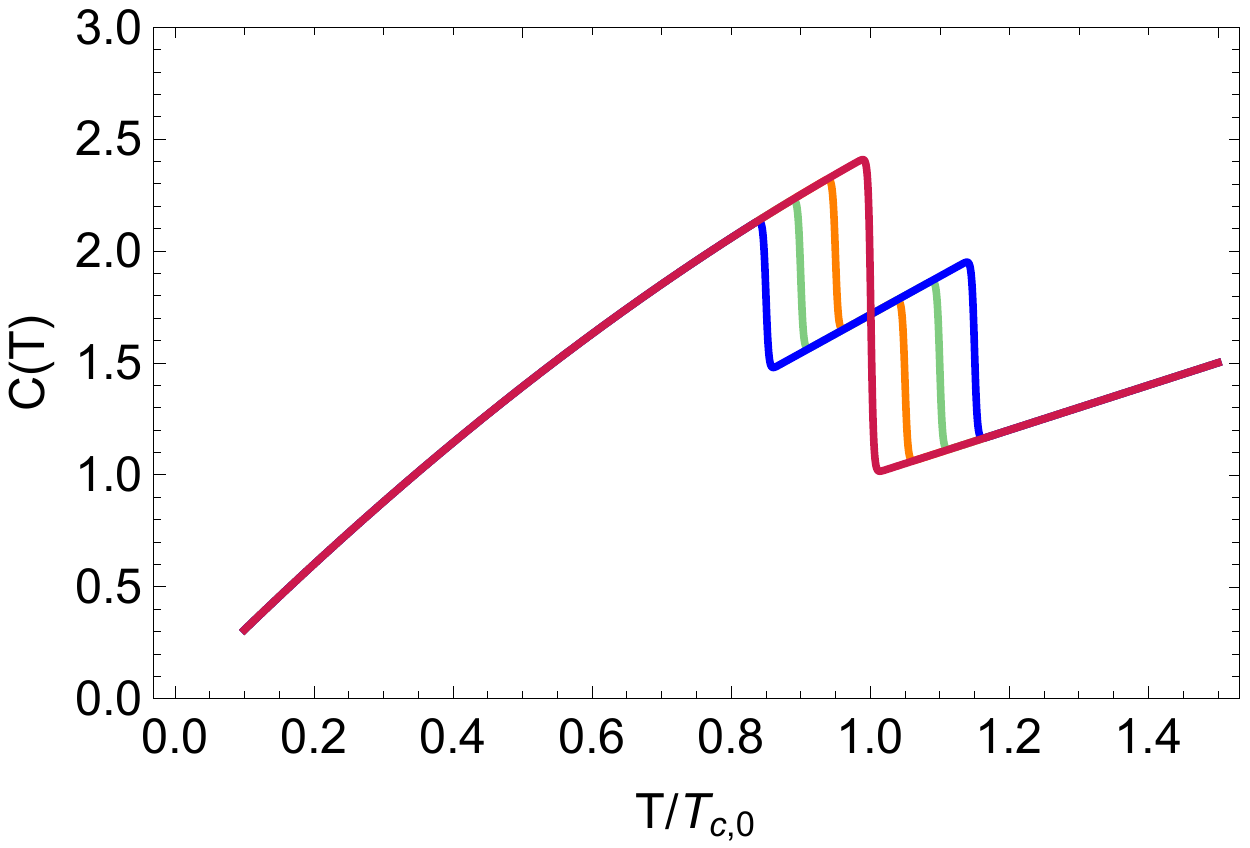}\includegraphics[scale=0.45]{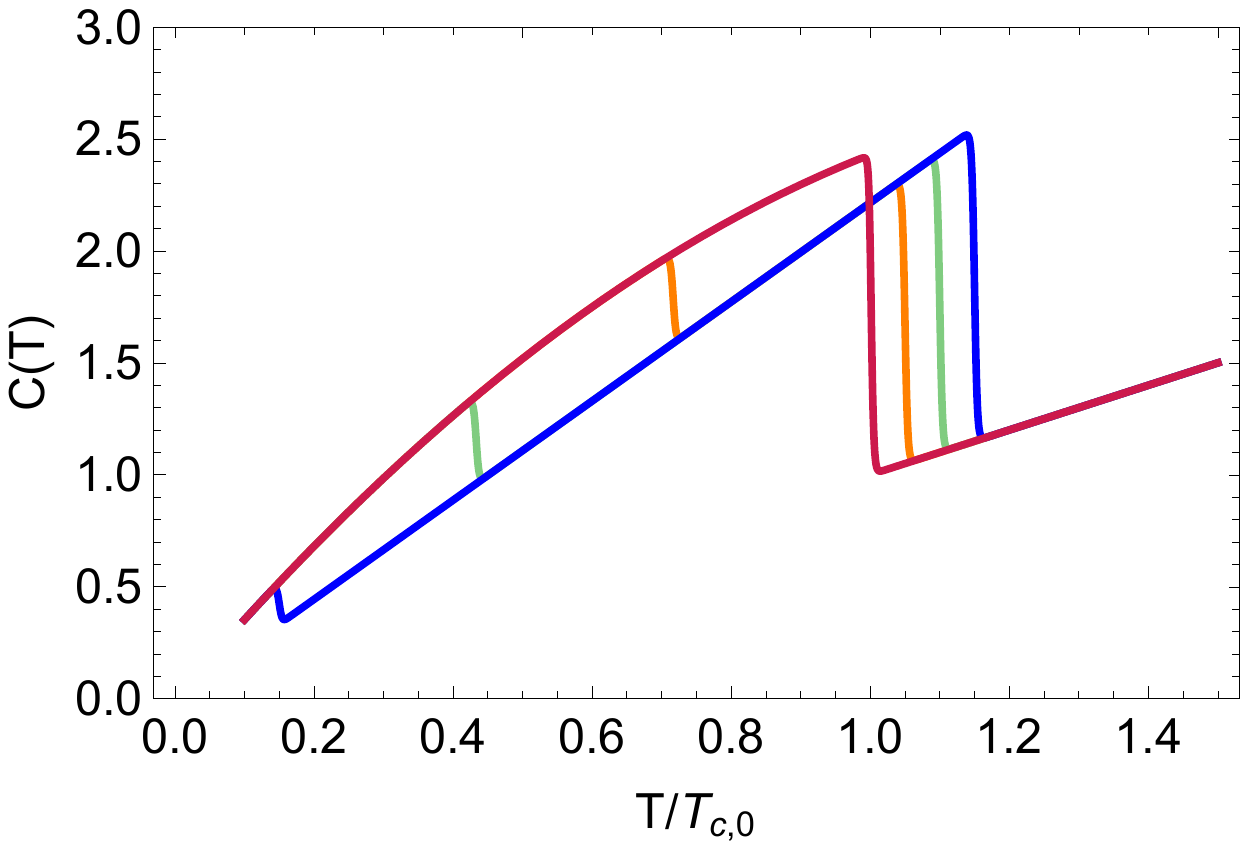}\includegraphics[scale=0.45]{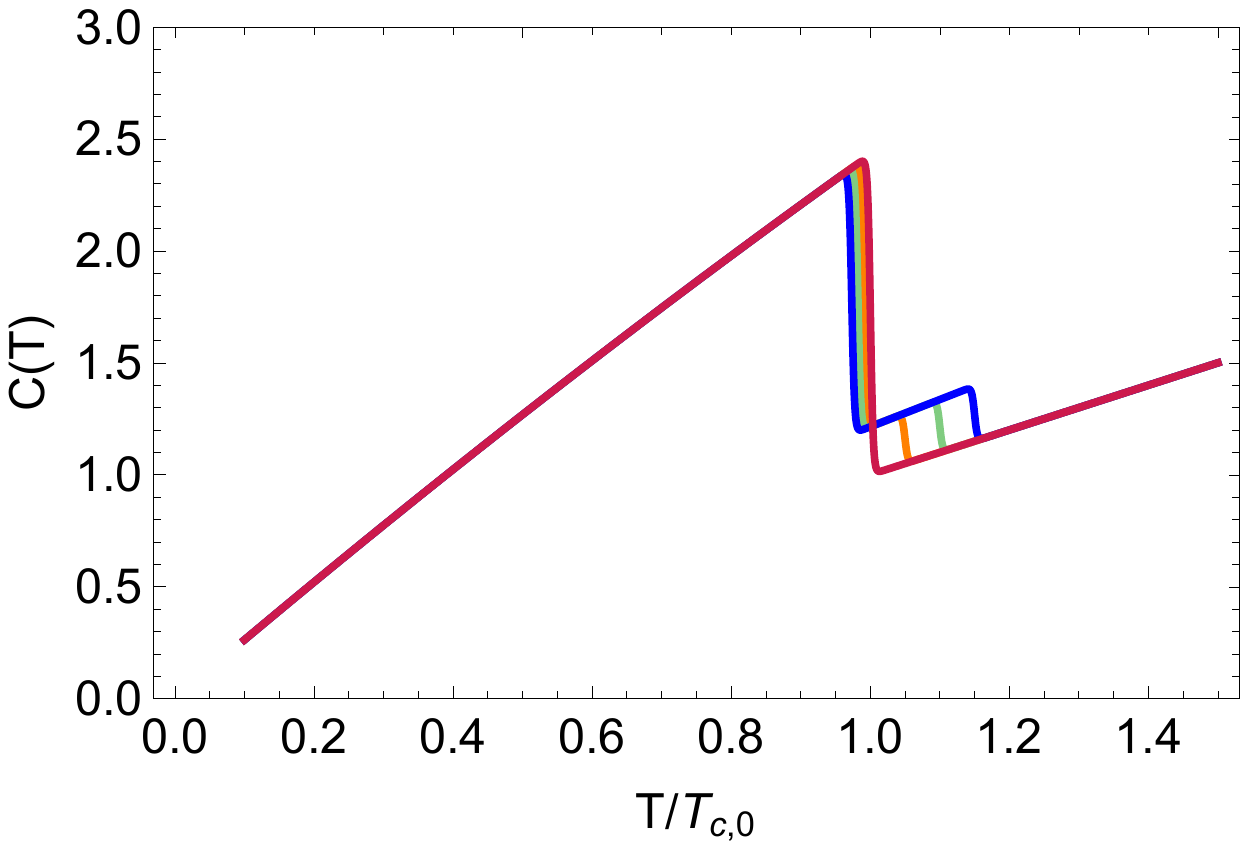}
\caption{Heat capacity as function of temperature for different strain values.
The left, middle and right panels correspond to the same interaction
parameters that give rise to the phase diagrams in Fig.~S\ref{Fig1}. Left panel:
For a symmetric splitting of the phase boundaries, the two heat capacity
jumps split symmetrically as well. Middle panel: a smaller jump at
the lower transition implies a rapid decrease of the phase boundary.
Right panel: the observed weak strain dependence of the lower transition
implies that the heat capacity jump at the lower transition dominates,
in disagreement with experiment.}\label{Fig2}
\end{figure*}

\begin{figure*}[tb!]
\includegraphics[scale=0.45]{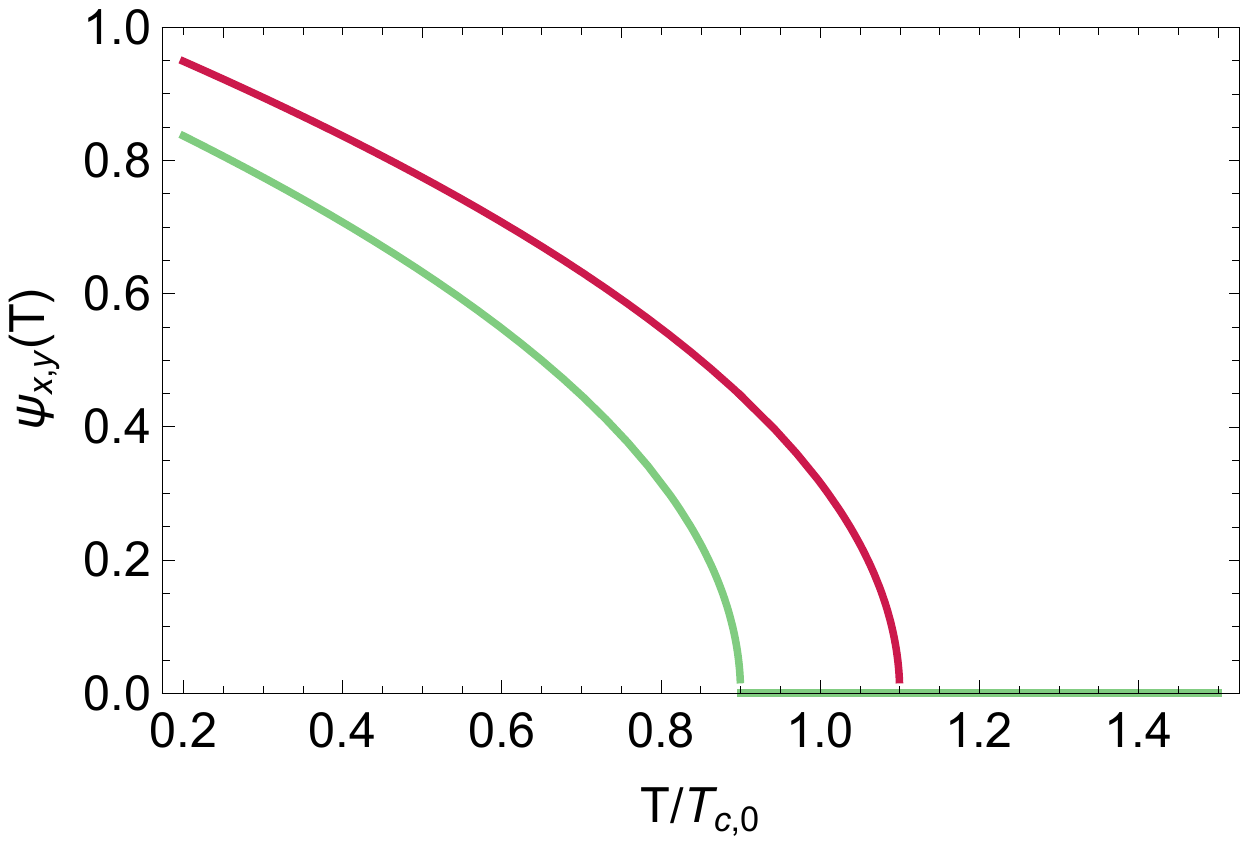}\includegraphics[scale=0.45]{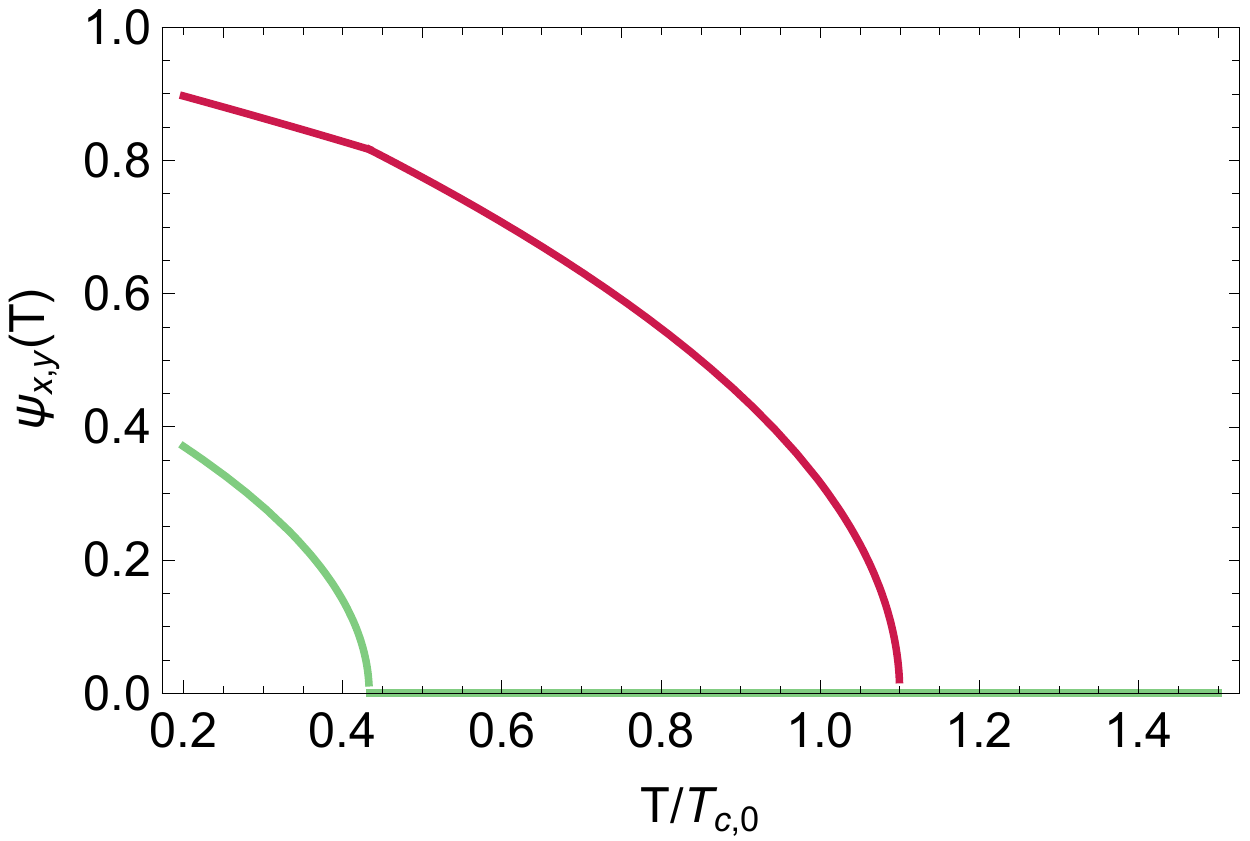}\includegraphics[scale=0.45]{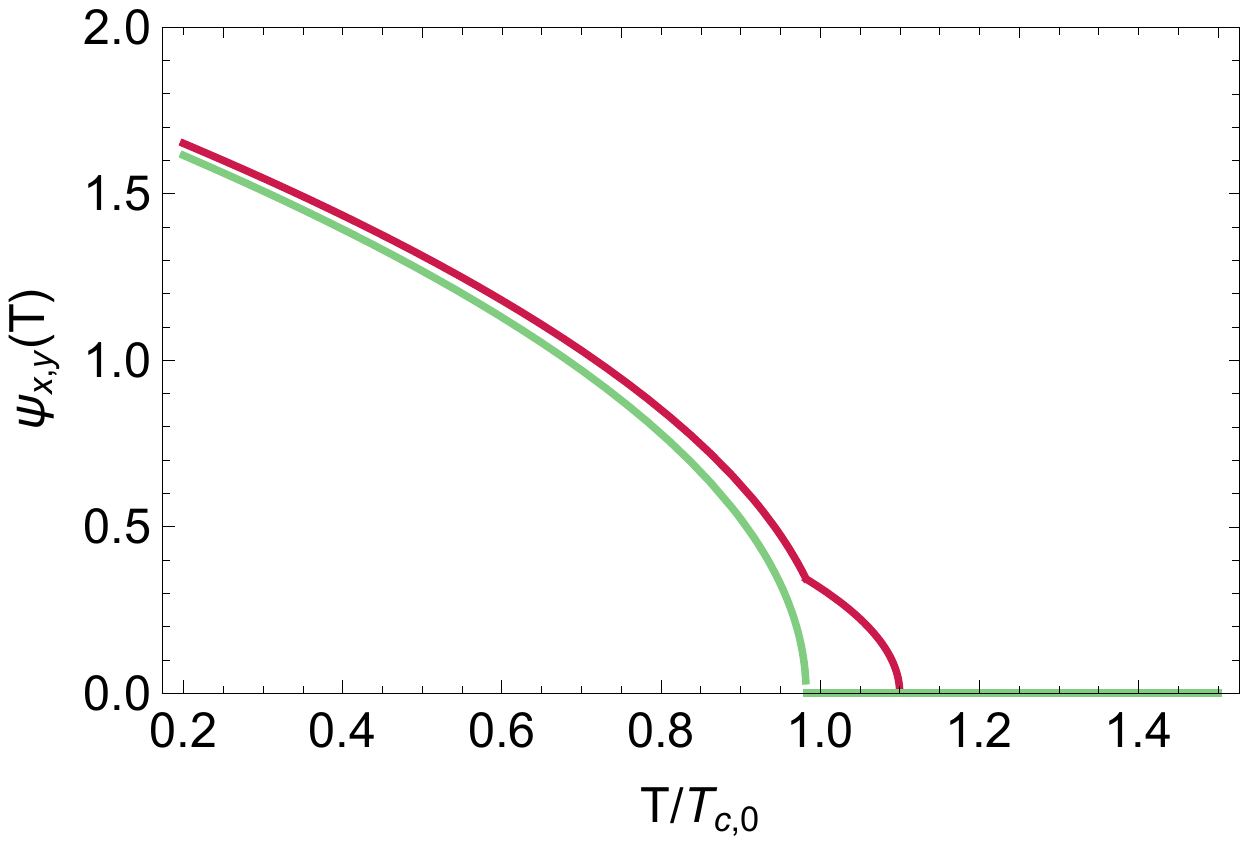}

\caption{Order parameters as function of temperature at finite strain for the
same interaction parameters that were used for the phase diagrams
in Fig.\ S\ref{Fig_phase_diagram}}

\label{Fig3}
\end{figure*}

This allows us to determine the specific heat $C=-T\frac{\partial^{2}F}{\partial T^{2}}$
with the result
\begin{equation}
C=C_{0}+\left\{ \begin{array}{ccc}
0 & {\rm ,if} & T_{c,u}<T\\
T\frac{a_{0}^{2}}{2u} & {\rm ,if} & T_{c,l}<T<T_{c,u}\\
T\frac{a_{0}^{2}}{u+w} & {\rm ,if} & T<T_{c,l}
\end{array}\right.
\end{equation}
Here $C_{0}\approx\gamma T+C_{{\rm latt}}$ is the normal state heat
capacity. Notice, for $\epsilon_{B_{1g}}=0,$ where we have only one transition,
the jump of the heat capacity at $T_{c,0}$ is $\Delta C_{0}=T_{c,0}\frac{a_{0}^{2}}{u+w}$.
Once the transition splits, the jump at the upper transition is
\begin{equation}
\frac{\Delta C\left(T_{c,u}\right)}{T_{c,u}}=\frac{a_{0}^{2}}{2u},
\end{equation}
while at the lower transition holds
\begin{eqnarray}
\frac{\Delta C\left(T_{c,l}\right)}{T_{c,l}} & = & a_{0}^{2}\left(\frac{1}{u+w}-\frac{1}{2u}\right).
\end{eqnarray}
The sum of the jumps is conserved, if compared
to the zero strain result:
\begin{equation}
\frac{\Delta C\left(T_{c,u}\right)}{T_{c,u}}+\frac{\Delta C\left(T_{c,l}\right)}{T_{c,l}}=\frac{\Delta C_{0}}{T_{c,0}}.
\end{equation}
Any deviation from this sum rule indicates strain effects that go
beyond the leading linear in $\epsilon_{B_{1g}}$ corrections. An example would
be the tuning to the van Hove point that is clearly an important higher
order effect that enters in strain dependencies of $a_{0},$ $u$,
etc.. If we assume
that $\eta<1$, i.e.\ that the strain dependence of the lower transition
is weaker than the dependence of the upper transition, we can use
$w<0$. This gives rise to the following inequality:
\begin{equation}
\frac{\Delta C\left(T_{c,l}\right)}{T_{c,l}}\geq\frac{a_{0}^{2}}{2u}=\frac{\Delta C\left(T_{c,u}\right)}{T_{c,u}}.
\end{equation}
Hence the jump divided by $T_{c}$ is then always larger at the lower
than at the upper transition. In particular, in the limit where $T_{c,l}$
is weakly strain dependent, we would  be in the regime where
$w\rightarrow-u$ such that instead $\frac{\Delta C\left(T_{c,l}\right)}{T_{c,l}}\gg\frac{\Delta C\left(T_{c,u}\right)}{T_{c,u}}$. In general it holds
\begin{equation}
\frac{\Delta C\left(T_{c,l}\right)}{T_{c,l}}=\frac{\Delta C\left(T_{c,u}\right)}{T_{cu}}\frac{\frac{dT_{c,u}}{d\epsilon_{B_{1g}}}}{\left|\frac{dT_{c.l}}{d\epsilon_{B_{1g}}}\right|}.
\end{equation}
This is illustrated in Fig.\ S\ref{Fig2}, where we show the corresponding
heat capacity anomalies for the phase diagrams that are summarized
in Fig.\ S\ref{Fig1}. The fact that the lower transition depends weakly
on strain would imply that the heat capacity anomaly at the lower
transition is the dominant one, in sharp contrast to the experimental
observation. For completeness, we also show in Fig.\ S\ref{Fig3} the
$T$-dependence of the order parameters at finite strain and for interaction
parameters that correspond to the three phase diagrams of Fig.\ S\ref{Fig1}.

\subsubsection*{{\boldmath $B_{2g}$}-nematic helical state}

$\left(1,\pm1\right)$ $B_{2g}$-nematic superconductor is the most
stable solution for $u'<0$ and $u''<0$, see Fig.\ S\ref{Fig_phase_diagram}.
Below the transition temperature $T_{c,0}$ the order parameter is
then given as
\begin{eqnarray}
\psi_{1} & = & e^{i\varphi}\sqrt{\frac{2a_{0}}{4u+u''}}\left(T_{c,0}-T\right)^{1/2},\nonumber \\
\psi_{2} & = & \pm \psi_{1} ,
\end{eqnarray}
 The energy of this state is $F=F_{0}-\frac{a^{2}}{4u+u''}$. Hence,
the heat capacity jump at the transition is $\Delta C\left(T_{c,0}\right)/T_{c,0}=\frac{2a^{2}}{4u+u''}$.

Under $B_{1g}$ strain the situation is in many ways analogous to the TRSB chiral state.
Let us assume without restriction $\lambda\epsilon_{B_{1g}}>0$. Above the
upper transition
\begin{equation}
T_{c,u}=T_{c,0}+\frac{\lambda}{a_{0}}\epsilon_{B_{1g}}
\end{equation}
it holds $\psi_{1}=\psi_{2}=0.$ In the intermediate temperature regime
$T_{c,l}<T<T_{c,u}$ holds
\begin{eqnarray}
\psi_{1} & = & e^{i\varphi}\sqrt{\frac{a_{0}}{u}}\left(T_{c,u}-T\right)^{1/2}\nonumber \\
\psi_{2} & = & 0
\end{eqnarray}
where the lower transition temperature
\begin{equation}
T_{c,l}=T_{c,0}-\eta\frac{\lambda}{a_{0}}\epsilon_{B_{1g}}
\end{equation}
is determined by a different dimensionless ratio of the interaction
parameters $\eta=\frac{4u+u''}{\left|u''\right|}>0$ such that $T_{c,l}<T_{c,0}$.
Below $T_{c,l}$ we have
\begin{eqnarray}
\psi_{1} & = & e^{i\varphi}\sqrt{\frac{2a_{0}}{4u+u''}}\left(T_{c,l}+2\eta\frac{\lambda\epsilon_{B_{1g}}}{a_{0}}-T\right)^{1/2}\nonumber \\
\psi_{2} & = & \pm e^{i\varphi}\sqrt{\frac{2a_{0}}{4u+u''}}\left(T_{c,l}-T\right)^{1/2}.
\end{eqnarray}
In particular this yields $\psi_{1}\left(T_{c,l}\right)=e^{i\varphi}\sqrt{\frac{4\lambda\epsilon_{B_{1g}}}{\left|u''\right|}}$.

Inserting these results for the order parameters into the free energy
yields
\begin{equation}
F=F_{0}+\left\{ \begin{array}{ccc}
0 & {\rm if} & T_{c,u}<T\\
-\frac{\left(a-\lambda\epsilon_{B_{1g}}\right)^{2}}{4u} & {\rm if} & T_{c,l}<T<T_{c,u}\\
-\frac{a^{2}}{4u+u''}+\frac{\lambda^{2}\epsilon_{B_{1g}}^{2}}{\left|u''\right|} & {\rm if} & T<T_{c,l}
\end{array}\right. \, .
\end{equation}
This allows us to determine the specific heat $C=-T\frac{\partial^{2}F}{\partial T^{2}}$
with the result
\begin{equation}
C=C_{0}+\left\{ \begin{array}{ccc}
0 & {\rm if} & T_{c,u}<T\\
T\frac{a_{0}^{2}}{2u} & {\rm if} & T_{c,l}<T<T_{c,u}\\
T\frac{2a_{0}^{2}}{4u+u''} & {\rm if} & T<T_{c,l}
\end{array}\right. \, .
\end{equation}
It follows for the heat capacity jump at the upper transition
\begin{equation}
\frac{\Delta C\left(T_{c,u}\right)}{T_{cu}}=\frac{a_{0}^{2}}{2u},
\end{equation}
while at the lower transition holds
\begin{eqnarray}
\frac{\Delta C\left(T_{c,l}\right)}{T_{cl}} & = & \frac{2a_{0}^{2}}{4u+u''}-\frac{a_{0}^{2}}{2u}\\
 & = & \frac{\Delta C\left(T_{c,u}\right)}{T_{cu}}\frac{\frac{dT_{c,u}}{d\epsilon_{B_{1g}}}}{\left|\frac{dT_{c.l}}{d\epsilon_{B_{1g}}}\right|}
\end{eqnarray}
and the ratios of the heat capacities are again given by the inverse
ratio of the corresponding changes in the transition temperatures.

\subsubsection*{{\boldmath $B_{1g}$}-nematic helical state}

The $\left(1,0\right)$, $\left(0,1\right)$ $B_{1g}$-nematic superconductor
is the most stable solution for $4u'<u''$ and $u''>0$,, see Fig.\ S\ref{Fig_phase_diagram}.
Below the transition temperature $T_{c,0}$ the order parameter is
then given as
\begin{eqnarray}
\psi_{1} & = & e^{i\varphi}\sqrt{\frac{a_{0}}{u}}\left(T_{c,0}-T\right)^{1/2},\nonumber \\
\psi_{2} & = & 0,
\end{eqnarray}
or $\psi_{1}\leftrightarrow\psi_{2}$. The energy of this state is
$F=F_{0}-\frac{a^{2}}{4u}$.

If ee include $B_{1g}$ strain a phase transition occurs at
\begin{equation}
T_{c}=T_{c,0}+\frac{1}{a_{0}}\left|\lambda\epsilon_{B_{1g}}\right|\, .
\end{equation}
If $\lambda\epsilon_{B_{1g}}>0$ the solution
\begin{eqnarray}
\psi_{1} & = & e^{i\varphi}\sqrt{\frac{a_{0}}{u}}\left(T_{c}-T\right)^{1/2},\nonumber \\
\psi_{2} & = & 0,
\end{eqnarray}
while changing the sign of $\lambda\epsilon_{B_{1g}}$ switches $\psi_{1}$
and $\psi_{2}$. In both cases the energy of the state is
\begin{equation}
F=F_{0}-\frac{\left(a-\left|\lambda\epsilon_{B_{1g}}\right|\right)^{2}}{4u}.\label{eq:Fnemlowest}
\end{equation}

In principle, there is another solution below the temperature $T^{*}=T_{c,0}-\frac{1}{a_{0}}\left|\lambda\epsilon_{B_{1g}}\right|$
which for $\lambda\epsilon_{B_{1g}}>0$ is given as
\begin{eqnarray}
\psi_{1}^{*} & = & 0,\nonumber \\
\psi_{2}^{*} & = & e^{i\varphi}\sqrt{\frac{a_{0}}{u}}\left(T^{*}-T\right)^{1/2}.
\end{eqnarray}
The energy of this other solution is $F^{*}=F_{0}-\frac{\left(a+\left|\lambda\epsilon_{B_{1g}}\right|\right)^{2}}{4u}$.
In the relevant temperature regime $T \leq T^{*}$ this energy is always
above $F$ given in Eq.\ \ref{eq:Fnemlowest}. Hence, this metastable
state is never the energetic lowest solution and we must use the free
energy of Eq.\ \ref{eq:Fnemlowest} for the heat capacity which is given
as
\begin{equation}
C\left(T\right)=C_{0}+\left\{ \begin{array}{ccc}
0 & \,\,{\rm if}\,\, & T>T_{c}\\
\frac{a_{0}^{2}}{2u}T & \,\,{\rm if}\,\, & T>T_{c}
\end{array}\right.,
\end{equation}
i.e.\ $\Delta C\left(T_{c}\right)/T_{c}=a_{0}^{2}/\left(2u\right)$
should be independent of strain.

\pnasbreak

\vspace{2em}

\bibliography{Sr2RuO4}

\end{document}